\documentclass[9pt,journal]{IEEEtran}

\usepackage{verbatim}
\usepackage{amsmath,amsthm,amsfonts,amssymb}
\usepackage{mathtools}
\usepackage{float}
\usepackage{color}
\usepackage{graphicx}

\usepackage{widetext}

\usepackage{bm}
\usepackage{balance}

\usepackage[caption=false,font=footnotesize]{subfig}
\usepackage{url}
\usepackage{upgreek}

\usepackage[switch]{lineno}
\usepackage[named]{algo}
\usepackage[noend]{algpseudocode}
\usepackage{algorithm}

\usepackage{cite}

\newtheorem{theorem}{Theorem}

\newcommand{\rr}{_\mathrm{r}}
\newcommand{\paren}[1]{\left({#1}\right)}
\newcommand{\bracket}[1]{{\left [{#1}\right ]}}

\begin{document}
\setlength{\abovedisplayskip}{3pt}
\setlength{\belowdisplayskip}{3pt}

\title{Displaced Sensor Automotive Radar Imaging}

\author{Guohua Wang and Kumar Vijay Mishra
\thanks{G. W. is with Hertzwell Pte Ltd, Singapore 138565, e-mail: guohua.wang@hertzwell.com.}
\thanks{K. V. M. is with the United States Army Research Laboratory, Adelphi, MD 20783 USA, e-mail: kumarvijay-mishra@uiowa.edu.}
\thanks{This work was funded by Hertzwell Pte Ltd.}
}

\maketitle

\begin{abstract}
Displaced automotive sensor imaging exploits joint processing of the data acquired from multiple radar units, each of which may have limited individual resources, to enhance the localization accuracy. Prior works either consider perfect synchronization among the sensors, employ single antenna radars, entail high processing cost, or lack performance analyses. Contrary to these works, we develop a displaced multiple-input multiple-output (MIMO) frequency-modulated continuous-wave (FMCW) radar signal model under coarse synchronization with only frame-level alignment. We derive Bayesian performance bounds for the common automotive radar processing modes such as point-cloud-based fusion as well as raw-signal-based non-coherent and coherent imaging. For the non-coherent mode, which offers a compromise between low computational load and improved localization, we exploit the block sparsity of range profiles for signal reconstruction to avoid direct computational imaging with massive data. For the high-resolution coherent imaging, we develop a method that automatically estimates the synchronization error and performs displaced radar imaging by exploiting sparsity-driven recovery models. Our extensive numerical experiments demonstrate these advantages. Our proposed non-coherent processing of displaced MIMO FMCW radars improves position estimation by an order over the conventional point-cloud fusion. 
\end{abstract}
\section{Introduction}
\label{sec:intro}
High-resolution sensing is a critical enabling technology for enhancing the safety of autonomous vehicles \cite{engels2017advances}. To this end, self-driving cars employ a number of sensors such as camera, radar, lidar, and ultrasonics to provide either individual or joint information from the surroundings \cite{patole2017automotive}. Although a camera is ideal for object detection and a lidar provides a very high range resolution, only a radar performs well in unfavorable conditions such as inclement weather and low visibility \cite{slavik2019phenomenological}. The comparatively low spatial resolution of a conventional radar is usually offset by by increasing its transmit signal bandwidth, coherent processing interval (CPI) or frame time \cite{richards2005fundamentals}, and antenna aperture size \cite{mishra2019toward}. However, limited frequency spectrum at millimeter-wave (mm-Wave) \cite{mishra2019toward,ayyar2019robust}, lower frame rate from increased frame times \cite{engels2017advances}, and requirement of small form factor restrict adoption of each of these measures \cite{meinel2014evolving,slavik2019cognitive,duggal2020doppler,wang2020stap}, respectively. In this context, deploying multiple radars on the same vehicle and then jointly processing their data to achieve high resolution has attracted significant attention within the automotive radar community \cite{steiner2019chirp,myakinkov2018distributed,khomchuk2016performance}. 

There is a large body of literature on distributed sensors for communications \cite{mudumbai2009distributed,tu2002coherent,chopra2016multistream} and radars \cite{tirer2016HRdpd,karlsson2017future,khomchuk2016performance}. Broadly these approaches for joint processing employ one of the two following two techniques. In \textit{geolocation database} method, cross-correlation of measurements of parameters such as directions-of-arrival (DoAs), time-differences-of-arrival (TDoAs), times-of-arrival (ToAs), and frequency-differences-of-arrival (FDoAs) of a radio-frequency (RF) signal received by multiple distributed sensors is used to retrieve the position of a target \cite{cong2002hybridTDOADOA,ho1993solution,chan1994simple}. The geolocation approach is easier to implement without any requirement of complex hardware. But it is inherently a two-step processing in which the measurement step is followed by position acquisition. Thus, the errors in each step propagate, thereby limiting the accuracy. This shortcoming is eliminated by employing \textit{direct position determination} (DPD) \cite{weiss2005direct,tirer2016HRdpd}, which infers geolocation directly from raw data. Nearly all of these methods assume that the distributed sensors are perfectly synchronized or their receiver clock offsets are known, which is impractical \cite{roehr2007method}. To alleviate this problem, some recent studies \cite{wang2018new,wang2017doa} developed procedures for source localization in communications by performing DoA estimation and DPD of multiple stationary RF transmitters without time synchronization. In the context of radar, \cite{berger2011noncoherent} suggested non-coherent processing using compressed sensing (CS) \cite{cho2018computable} to estimate target positions for widely distributed radars without synchronization. However, it exploits only range information and is inferior to techniques that additionally use other parameters such as DoA for localization.

Further, geolocation approaches do not yield accurate estimates of target reflectivity. This has led to the development of \textit{distributed imaging} algorithms which yield both target position and reflectivity. Recently, \cite{liu2015sparsity} proposed a CS-based high-resolution multi-static radar imaging using raw data in spectral domain; perfect synchronization and perfect knowledge of sensor geometry was assumed. More recently, an interesting study \cite{lodhi2019coherent} on coherent radar imaging using unsynchronized distributed antennas modeled errors in time synchronization and antenna positions to accurately estimate target reflectivity and position. 
However, this work did not provide any theoretical guarantees. Further, for automotive radars, the assumption on bounded time synchronization errors is impractical \cite{roehr2007method}. In this paper, contrary to prior works, we derive theoretical performance limits of displaced radar imaging, assume imperfect synchronization, and apply automotive-specific system details.


In our model, different from a conventional distributed multiple-input multiple-output (MIMO) radar \cite{sun2019target}, the automotive displaced sensors operate independently 
and are only \textit{coarsely} time-synchronized through use of standards such as IEEE 1588 generic precision time protocol (gPTP) \cite{ieee2008precision}, network time protocol (NTP) \cite{mills1991internet} and wireless PTP \cite{garg2017wireless}. These cost-effective clock synchronization protocols are also popular in other applications, including electrical grid networks, cellular base-station synchronization, and industrial control \cite{karthik2020robust}. While their accuracy is comparable to Global Positioning System (GPS)-based timing of microseconds, the resulting synchronization is coarser than conventional mm-Wave TDoA-based localization and positioning. 
Each radar being independent, the received signal depends on only local timing of each sensor thereby circumventing the need of fine inter-sensor time-synchronization. 

In practice, modern vehicles are fitted with sensors so that any displacements in radar positions are insignificant. Therefore, in this work, we consider errors arising from solely the coarse synchronization among radar sensors. We first study the performance bounds on the imaging accuracy for different common automotive radar processing modes, i.e., point-cloud fusion, non-coherent imaging and coherent imaging. 
Since prior information from an initial imaging of the target environment and high definition maps are readily available for automotive radars, we adopt the Bayesian approach to derive the error bounds. Note that, among prior works, \cite{khomchuk2016performance} derived deterministic bounds for the case of stationary widely separated MIMO radars in non-automotive applications. A single moving colocated MIMO radar was considered in \cite{boyer2010performance}. 

Our analysis shows that the non-coherent and coherent imaging exhibit better localization than the point-cloud fusion. Henceforth, in this work, we focus on developing imaging algorithms only for non-coherent and coherent cases. We formulate the non-coherent processing as a block-sparse recovery problem in a reduced-rate sensing framework \cite{mishra2019sub}. Other recent works on single-sensor automotive \cite{simoni2019height} and MIMO imaging \cite{hu2018mimo} harness block-sparsity of range profiles to mitigate the processing problem with massive data samples. Our approach exploits block sparsity across profiles from multiple sensors that are not perfectly synchronized. Although computationally efficient, this non-coherent processing only provides a limited improvement in resolution. Therefore, we further devise a high-resolution coherent imaging, with online correction for the time synchronization error, which is based on both conventional compressed sensing and Bayesian sparse reconstruction. Preliminary results of this work appeared in our conference publication \cite{wang2020performance}, which did not include multiple targets, actual imaging performance, reconstruction algorithms, and comparison of various modes. 



 The rest of the paper is organized as follows. In the next section, we describe the coarse synchronization signal model for the conventional frequency-modulated continuous-wave (FMCW) MIMO radar operating in time-division multiplexing (TDM) mode. 
 For this system, we introduce the aforementioned three processing modes relevant to the automotive radar imaging. Then, in Section~\ref{sec:perf_bounds}, we derive the Bayesian Cram\'{e}r-Rao lower bound (BCRLB)\footnote{Note that BCRLB is different from the hybrid CRLB (HCRLB) \cite{rockah1987array,noam2009notes}, where parameter vector has both random and deterministic variables.} \cite{kumar2018information,mishra2017performance} for estimating the position in these modes. In Section~\ref{sec:displ}, we develop our imaging algorithms for non-coherent and coherent processing. We validate our model and methods with extensive numerical experiments in Section~\ref{sec:numexp} before concluding in Section~\ref{sec:summ}. 
 
 Throughout this paper, we denote boldface lowercase, boldface uppercase and calligraphic letters for vectors, matrices and index sets, respectively. The notation $(\cdot)^H$ stands for conjugate transpose and transpose for complex and real quantities, respectively. The Kronecker and Hadamard products are written as $\otimes$ and $\circ$, respectively. We use $\mathbf{I}_N$ for the identity matrix of size $N \times N$. The functions $\Re(\cdot)$ and $\Im(\cdot)$ yield the real and imaginary parts of their arguments; $\mathrm{diag}\{\mathbf{a}\}$ is a diagonal matrix formed from the elements of vector $\mathbf{a}$; $\textrm{vec}(\cdot)$ vectorizes the matrix argument column-wise; $\circ(\cdot)$ is an upper bound that cannot be tight; and $\mathbb{E}\{\cdot\}$ is the statistical expectation.
\section{System Model}
\label{sec:sys_mod}
\begin{figure}[t]
    \centering
    \includegraphics[width=1.0\columnwidth]{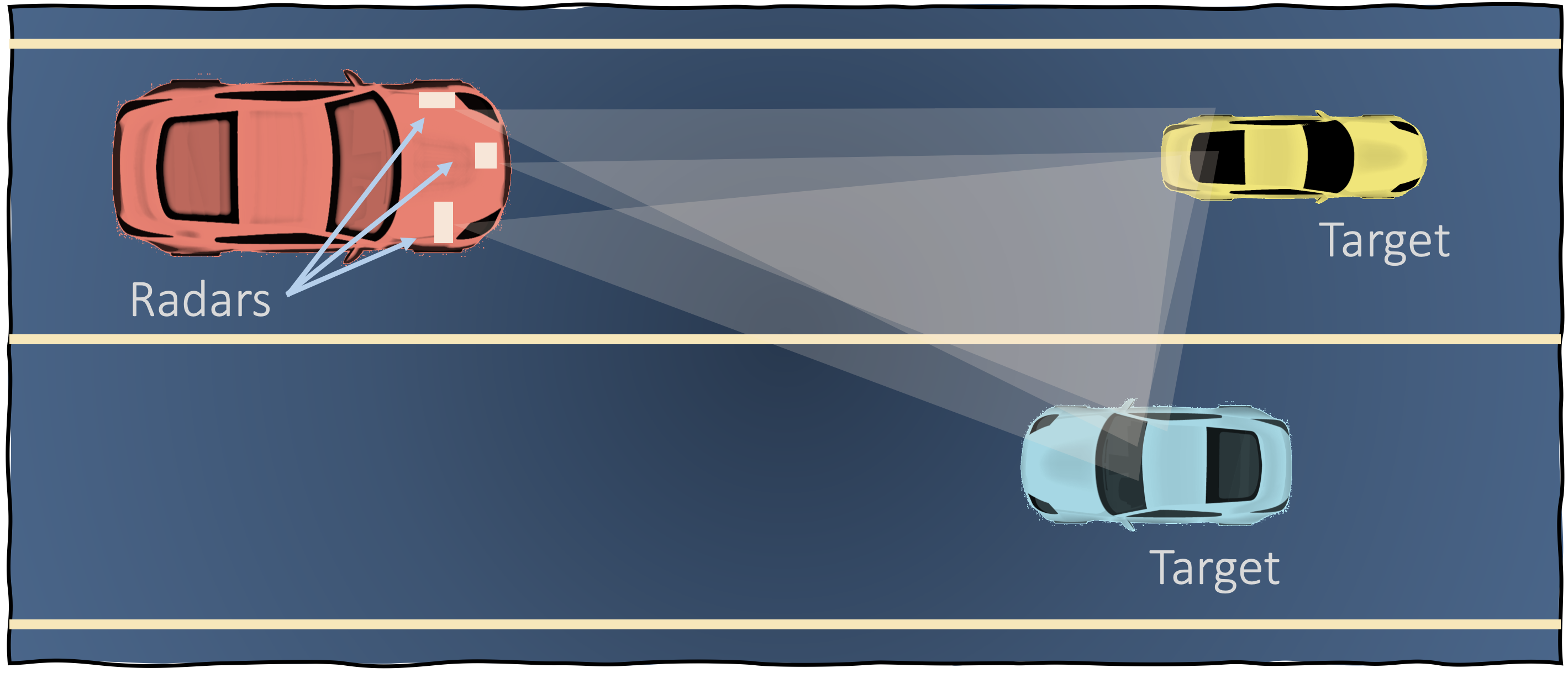}
    \caption{An automotive displaced sensor system employs multiple radars (pink rectangles) that are mounted on the different locations of the same vehicle (orange) and jointly observe common targets (blue and yellow vehicles). The targets may also be stationary objects on the road and urban infrastructure.}
    \label{fig:radarconfig}
\end{figure}

Consider a displaced automotive radar system (Fig.~\ref{fig:radarconfig}) with $Q$ sensors mounted on different locations of a single vehicle. Without loss of generality, each radar is a MIMO array with $N$ transmit and $M$ receive antennas. 
The center-of-mass of the vehicle is the global reference position (or origin) across this system. Each radar sensor has some error in its position, $\mathbf{p}^q_e=[x_{e,q};y_{e,q};z_{e,q}]$. 
In a three-dimensional (3-D) coordinate system, positions of transmit and receive antennas of $q$-th radar are\par\noindent\small
\begin{align}
\tilde{\mathbf{p}}^q_{T,n}=[x_{q,n};y_{q,n};z_{q,n}]+\mathbf{p}^q_e=\mathbf{p}^q_{T,n}+\mathbf{p}^q_e\in \mathbb{R}^{3\times 1}, n=1,\cdots,N,
\end{align}\normalsize
and \par\noindent\small
\begin{align}
\tilde{\mathbf{p}}^q_{R,m}=[x_{q,m};y_{q,m};z_{q,m}]+\mathbf{p}^q_e=\mathbf{p}^q_{R,m}+\mathbf{p}^q_e\in \mathbb{R}^{3\times 1} , m=1,\cdots,M,
\end{align}\normalsize
respectively. All radars move along with the vehicle at an identical speed of $\mathbf{v}=[v_x,v_y,v_z]^T \in \mathbb{R}^{3 \times 1}$. The radar signals are transmitted in TDM mode, which is a widely adopted waveform orthogonality in automotive MIMO radars \cite{patole2017automotive,wang2020stap}. 
The cross-interference between individual radar sensors is avoided by separating the transmit spectrum of each radar through frequency diversity \cite{richards2005fundamentals}. 

Each $q$-th radar transmitter emits $K$ FMCW chirps, each of duration $T_p$, at a pulse repetition interval $T_r$ and carrier frequency $f_c^q$  and modulation rate $B_r$. The frame time (or CPI) comprises $NK$ sweeps from all transmitters. 
The proposed system here is synchronized only coarsely through gPTP or similar protocols. The coarse clock implies that, for a global time reference $t$, the time offset with the local time $t_q$ of $q$-th radar is $\sigma_q$, so that $t_q=t-\sigma_q$. The $\sigma_q$ is very small, usually of the order of milliseconds. As a consequence, the radar and target positions are assumed to be constant across all different radar sensors during a CPI \cite{mishra2019toward}. 
The transmit waveform at $k$-th pulse and $n$-th transmit antenna of the $q$-th radar sensor is\par\noindent\small
\begin{align}
    s_{q,n}(t,k)&=\textrm{rect}\left(\frac{t-\sigma_q}{T_p}\right)\nonumber\\
    & e^{\mathrm{j}2\pi f_c^q(t-\sigma_q+(n-1+(k-1)N)T_r)}e^{\mathrm{j}\pi B_r(t-\sigma_q)^2}, \nonumber\\
    & k=0,\cdots, K ,
\end{align}\normalsize
where the rectangular pulse\par\noindent\small
\begin{align}
    \textrm{rect}(t)=\begin{dcases}
         & 1, \; t\in [0,T_p] \\
         & 0, \;\textrm{otherwise}.
    \end{dcases}
\end{align}\normalsize

The transmit chirps impinge a target at position $\mathbf{p}_t=[x;y;z]\in \mathbb{R}^{3\times 1}$. The target Doppler velocity is $
    v_q=\mathbf{v}_c^H\mathbf{p}_{t,q}$,
where 
$\mathbf{p}_t^q$ denotes the direction vector between the $q$-th radar and target. The target is relatively far from the radar sensor as compared to the aperture of each radar sensor. Thus, the direction vector, constant for each radar sensor, is  $\mathbf{p}_{t,q}=[\cos(\theta_q)\sin(\phi_q);\sin(\theta_q)\sin(\phi_q);\cos(\phi_q)]$, where $\theta_q$ and $\phi_q$ stand for the azimuth and elevation of target as viewed from the $q$-th radar sensor, respectively. The bistatic time delay from the $n$-th transmitter to target and back to the $m$-th receiver of the $q$-th radar sensor is\par\noindent\small
\begin{align}
    \tau_{q,m,n}(t,k)=\frac{\tilde{g}_{q,m,n}}{c}+2v_q(t+(n-1+(k-1)N)T_r)/c,
\end{align}\normalsize
where the bistatic range $\tilde{g}_{q,m,n}=\tilde{g}^q_{T,n}+\tilde{g}^q_{R,m}$
 with \par\noindent\small
 \begin{equation}
\begin{split}
    &\tilde{g}^q_{T,n}=||\mathbf{p}_t-\tilde{\mathbf{p}}^q_{T,n}||\\
    &=((x_{q,n}+x_{e,q}-x)^2+(y_{q,n}+y_{e,q}-y)^2 +(z_{q,n}+z_{e,q}-y)^2)^{1/2},  \end{split}
 \end{equation}\normalsize
 and\par\small\noindent
 \begin{equation}
 \begin{split}
    &\tilde{g}^q_{R,m} =||\mathbf{p}_t-\tilde{\mathbf{p}}^q_{R,m}|| \\
    &=
    ((x_{q,m}+x_{e,q}-x)^2+(y_{q,m}+y_{e,q}-y)^2 +(z_{q,m}+z_{e,q}-y)^2)^{1/2}.
\end{split}
\end{equation}\normalsize

Note that the bistatic range depends on the position error of each radar sensor. In practice, modern vehicle manufacturing allows control of sensor position error to sub-centimeter levels \cite{leitner2019validation}. Hence, considering this error is much smaller than the range resolution and antenna array aperture, applying Taylor series expansion yields \par\noindent\small
\begin{equation}
\begin{split}
    \tilde{g}^q_{T,n}&=||\mathbf{p}_t-\tilde{\mathbf{p}}^q_{T,n}||=||\mathbf{p}_t-\mathbf{p}^q_{T,n}|| +\mathbf{p}^H_{e,q}\mathbf{p}_{t,q}=g^q_{T,n}+\circ(g^q_{T,n}),
 \end{split}
 \end{equation}\normalsize
and \par\noindent\small
\begin{equation}
\begin{split}
    \tilde{g}^q_{R,m}&=||\mathbf{p}_t-\tilde{\mathbf{p}}^q_{R,m}||=||\mathbf{p}_t-\mathbf{p}^q_{R,m}|| +\mathbf{p}^H_{e,q}\mathbf{p}_{t,q}=g^q_{R,m}+\circ(g^q_{R,m}).
 \end{split}
 \end{equation}\normalsize
 The small position error of each radar sensor, hence, does not affect the range and DoA estimates of targets. 
 The bistatic range becomes\par\noindent\small
 \begin{equation}
     \tilde{g}_{q,m,n}=g_{q,m,n}+\circ_q,
     \label{eq:bisR}
 \end{equation}\normalsize
 where $g_{q,m,n}=g^q_{T,n}+g^q_{R,m}$ and $\circ_q=\circ(g_{q,m,n})$. The target follows the Swerling I model \cite{skolnik2008radar} so that its unknown reflection coefficient $\tilde{\alpha}_q$ remains constant across the CPI. Only when the view angles from different radar sensors are significantly different, the target coefficient is considered different for each radar. This is the case with non-coherent imaging. In coherent processing, the target reflectivity is identical across all sensors.

The signal reflected off the target and received at $m$-th antenna is\par\noindent\small
\begin{flalign}
    &s_{q,m,n}(t,k)=\tilde{\alpha}_q\textrm{rect}\left(\frac{t-\sigma_q-\tau_{q,m,n}(t,k)}{T_p}\right) \nonumber\\
    & e^{\mathrm{j}2\pi f_c^q(t-\sigma_q-\tau_{q,m,n}(t,k)+(n-1+(k-1)N)T_r)}  e^{\mathrm{j}\pi B_r(t-\sigma_q-\tau_{q,m,n}(t,k))^2}.
\end{flalign}\normalsize
The FMCW receiver mixes this signal with the transmit waveform of the same radar transmitter to produce the baseband signal\par\noindent\small
\begin{flalign}
  &\tilde{y}_{q,m,n}(t,k) = \tilde{\alpha}_q\textrm{rect}\left(\frac{t-\sigma_q-\tau_{q,m,n}(t,k)}{T_p}\right) \nonumber\\ 
  & e^{-\mathrm{j}2\pi f_c^q\tau_{q,m,n}(t,k)} e^{\mathrm{j}\pi B_r(-2(t-\sigma_q)\tau_{q,m,n}(t)+\tau_{q,m,n}^2(t,k))}.
\end{flalign}\normalsize
Changing the variables $t=t_c+\sigma_q$ yields\par\noindent\small
\begin{flalign}
  &\tilde{y}_{q,m,n}(t_c,k) = \tilde{\alpha}_q\textrm{rect}\left(\frac{t_c-\tau_{q,m,n}(t_c+\sigma_q,k)}{T_p}\right) \nonumber\\ 
  & e^{-\mathrm{j}2\pi f_c^q\tau_{q,m,n}(t_c+\sigma_q,k)} e^{\mathrm{j}\pi B_r(-2(t_c)\tau_{q,m,n}(t_c+\sigma_q,k)+\tau_{q,m,n}^2(t_c+\sigma_q,k))}.
\end{flalign}\normalsize
Substituting delay expression and omitting higher order phases gives \par\noindent\small
\begin{align}
    y_{q,m,n}(t_c,k)&=\tilde{\alpha}_q c_{q}c_{\circ,q} e^{-\mathrm{j}2\pi f_c^qg_{q,m,n}/c} \nonumber\\
    & e^{-\mathrm{j}2\pi\left(\frac{2f_c^qv}{c}+\frac{B_rg_{qmn}}{c}\right)t_c}e^{-\mathrm{j}2\pi\frac{2f_c^qvT_r}{c}(n-1+(k-1)N))},
\label{eq:sqmntk}
\end{align}\normalsize
where $c_q=e^{-\mathrm{j}2\pi f_c^q\frac{2v_q}{c}\sigma_q}$ is the complex phase term due to $\sigma_q$ and $c_{\circ,q}=e^{-\mathrm{j}2\pi\frac{B_r \circ_q}{c}t_c}$ is the phase term due to sensor position error. 

By definition, we note that both $c_q$ and $c_{o,q}$ are dependent on the positions of target and $q$-th radar sensor. This makes the phase uncertainty different from the conventional phase errors encountered in autofocusing problems of synthetic aperture radar (SAR) imaging \cite{koo2005comparison}. Owing to the small position error in mounting the radars, we omit the high-order term $o_q$ in the sequel. Thus, (\ref{eq:sqmntk}) becomes\par\noindent\small
\begin{align}
    y_{q,m,n}(t_c,k)&=\tilde{\alpha}_q c_{q} e^{-\mathrm{j}2\pi f_c^qg_{q,m,n}/c} \nonumber\\
    & e^{-\mathrm{j}2\pi\left(\frac{2f_c^qv}{c}+\frac{B_rg_{qmn}}{c}\right)t_c}e^{-\mathrm{j}2\pi\frac{2f_c^qvT_r}{c}(n-1+(k-1)N))}.
\label{eq:sqmntkv0}
\end{align}\normalsize
Then, incorporating the time-synchronization-induced phase term $c_q$ into the signal amplitude, i.e., $\alpha_q=\tilde{\alpha}_qc_q$, gives\par\noindent\small
\begin{align}
    y_{q,m,n}(t_c,k)&=\alpha_q e^{-\mathrm{j}2\pi f_c^qg_{q,m,n}/c} \nonumber\\
    & e^{-\mathrm{j}2\pi\left(\frac{2f_c^qv}{c}+\frac{B_rg_{qmn}}{c}\right)t_c}e^{-\mathrm{j}2\pi\frac{2f_c^qvT_r}{c}(n-1+(k-1)N))}.
\label{eq:sqmntkv1}
\end{align}\normalsize

In (\ref{eq:sqmntkv1}), we used $t_c$ to replace $t_q$ because the relative local timing at different sensors is same. This implies that when each radar sensor is synchronized within itself, the received signal model - except for a complex phase term - depends on only local time, which is same for all radars. This holds as long as the frame times of different radars are coarsely synchronized. In addition, the above calculations reveal that the time synchronization primarily affects the Doppler frequency estimates. For static target and radar, the model is identical for both synchronous and asynchronous operations. In automotive scenarios, radars often encounter several stationary targets on and alongside the road. 
The static target model 
simplified as\par\noindent\small
\begin{align}
    s_{q,m,n}(t_c)=\alpha_q Ke^{-\mathrm{j}2\pi (f_c^q+\frac{B_r}{c}t_c)g_{q,m,n}/c}
    \triangleq \alpha_q h_{qmn}(t_c).
    \label{eq:sqmnStatic}
\end{align}\normalsize

\section{Joint Received Signal Processing}
\label{sec:joint_rx}
We now consider three different automotive radar techniques to jointly process the received signal. The first method employs radar point clouds that are the result of the received signal processed by constant false-alarm rate (CFAR) algorithms, which return those reflection points that exceed reflection amplitude thresholds. The non-coherent and coherent imaging methods operate on raw signals (instead of the processed estimates used by point-cloud fusion). 
\subsection{Point-cloud fusion} 
A point-cloud generated at each $q$-th radar sensor is a set of 4-D points that comprise estimates of target's range, DoA, Doppler, and amplitude as $ \{\hat{r}_q,\hat{\theta}_q,\hat{\phi}_q,\hat{v}_q, \hat{\alpha}_q\}$. The estimation accuracy of these parameters is improved by fusing the point-clouds of each sensor. 
The displaced sensor imaging refines target estimation based on the set of measurements from all sensors. For example, consider the estimation of target position. Then, point-cloud fusion exploits both $\hat{r}_q$, and $\hat{\theta}_q$ and $\hat{\phi}_q$ as follows. Define $\bm{\Phi}=[x,y,z]^T$ as unknown parameter vector of target position. The estimates of range and DoA are\par\noindent\small
\begin{align}
    \hat{r}_q&=r_q(\bm{\Phi})+n_{r,q},\nonumber\\
    \hat{\theta}_q&=\theta_q(\bm{\Phi})+n_{\theta,q},\nonumber\\
    \hat{\phi}_q&=\phi_q(\Phi)+n_{\phi,q}
\end{align}\normalsize
where $n_{r,q}$, and $n_{\theta,q}$ and $n_{\phi,q}$ are the measurement noises. The range and DoA measurements at the $q$-th radar are, respectively,\par\noindent\small
\begin{align}
    r_q(\bm{\Phi})&=((x-x_{q,1})^2+(y-y_{q,1})^2)^{1/2},\\
    \theta _q(\bm{\Phi})&=tan^{-1}\left(\frac{y-y_{q,1}}{x-x_{q,1}}\right),\\
    \phi_q(\bm{\Phi})&=tan^{-1}\left(\frac{z-z_{q,1}}{((x-x_{q,1})^2+(y-y_{q,1})^2)^{1/2}}\right).
\end{align}\normalsize
The point-cloud fusion does require the position of individual radar sensor which may affect the performance. However, as mentioned earlier, this position error can be ignored in the fusion processing. 
Stacking measurements from all $Q$ radars gives\par\noindent\small
\begin{equation}
    \mathbf{z}=\mathbf{f}(\bm{\Phi})+\mathbf{n},
\end{equation}\normalsize
where $\mathbf{z}=[\hat{r}_1,\cdots,\hat{r}_Q,\hat{\theta}_1,\cdots,\hat{\theta}_Q]^T$, $\mathbf{f}(\bm{\Phi})=[ r_1(\bm{\Phi}), \cdots, r_Q(\bm{\Phi}),$ $\theta_1(\bm{\Phi}), \cdots, \theta_Q(\bm{\Phi}) ]^T$, and $\mathbf{n}\sim N(\mathbf{0},\mathbf{R}_n)$ is Gaussian noise. The probability density function of the measurements $\mathbf{z}$ is $p(\mathbf{z},\bm{\Phi})= p(\mathbf{z}|\bm{\Phi})p_o(\bm{\Phi})$. The point-cloud fusion aims at improving the accuracy of the estimation of $\bm{\Phi}$ by maximizing the posterior distribution.
\subsection{Non-coherent processing} 
In non-coherent case, the target reflection coefficient observed at each radar sensor is different. Per (\ref{eq:sqmntkv1}), in the presence of noise, the $n_s=1,\cdots,N_s$ discrete-time non-coherent measurements after sampling at interval $T_s$ in fast-time are\par\noindent\small
\begin{equation}
    z_{q,m,n}(n_s,k)=\alpha_q h_{q,m,n}(n_s,k)+n_{q,m,n}(n_s,k),
\end{equation}\normalsize
where \par\noindent\small
\begin{equation}
    \begin{split}
    h_{q,m,n}(n_s,k)&=e^{-\mathrm{j}2\pi f_c^qg_{q,m,n}/c}\\
    & e^{-\mathrm{j}2\pi\left(\frac{2f_c^qv}{c}+\frac{B_rg_{qmn}}{c}\right)n_sT_s}e^{-\mathrm{j}2\pi\frac{2f_c^qvT_r}{c}(n-1+(k-1)N))}.
    \end{split}
    \label{eq:hqmnnsk}
\end{equation}\normalsize
Stacking these in a single vector, we obtain $\mathbf{z}_{nc}=[z_{1,1,1}(1,1),\cdots,z_{Q,M,N}(N_s,K)]^T$ such that\par\noindent\small
\begin{align}
    \mathbf{z}_{nc}=(\bm{\alpha}\otimes I_{MNKN_s\times 1})\circ\mathbf{h}(\bm{\Phi})+\mathbf{n}_{nc},
    \label{eq:RxSigModNCP}
\end{align}\normalsize
where $\bm{\alpha}=[\alpha_1,\cdots,\alpha_Q]^T$ and
$\mathbf{h}(\bm{\Phi})=[h_{111}(1,1),\cdots,$ $h_{QMN}(N_s,K)]$. We further define $\bm{\Phi}_1=[\bm{\Phi}^T,\bm{\alpha}]^T$. The noise follows circular symmetric Gaussian distribution, i.e., $\mathbf{n}_{nc}\sim \mathcal{CN}(\mathbf{0},\mathbf{R}_{nc})$ and $\alpha_q$ are i.i.d Gaussian random variables following $\alpha_q \sim \mathcal{CN}(0,\sigma^2)$.
The probability density function is conditioned on the amplitude as $p(\mathbf{z}_{nc},\bm{\Phi}|\bm{\alpha})= p(\mathbf{z}_{nc}|\bm{\Phi},\bm{\alpha})p_o(\bm{\Phi})$, where $\bm{\Phi}$ is independent of $\bm{\alpha}$, i.e., $p_o(\bm{\Phi})=p_o(\bm{\Phi}|\bm{\alpha})$.
The non-coherent processing method estimates the target position based on the posterior probability distribution.

\subsection{Coherent processing} 
When the antennas are well-calibrated over the sensors, then all radars view the target with identical reflection coefficient, i.e., $\tilde{\alpha}_q=\alpha$. Conventional coherent processing assumes perfect synchronization so that $c_q$ in (\ref{eq:sqmntkv1}) is irrelevant here. These are strong and often impractical assumptions for automotive applications. However, coherent processing does provide the best achievable performance bound and serves as a benchmark. In practice, the antennas could be calibrated in advance and as long as the synchronization error is estimated correctly, coherent processing could still be employed. The measurements for the coherent processing are\par\noindent\small
\begin{equation}
    \mathbf{x}_c=\alpha \mathbf{h}(\bm{\Phi})+\mathbf{n}_c,
\end{equation}\normalsize
where the noise $\mathbf{n}_c\sim \mathcal{CN}(\mathbf{0},\mathbf{R}_{c})$ and $\alpha \sim \mathcal{CN}(0,\sigma^2)$. Define $\bm{\Phi}_2=[\bm{\Phi}^T, \alpha]^T$. Assuming the signal parameter $\alpha$ is independent of $\bm{\Phi}$, the probability density function is $p(\mathbf{x}_c,\bm{\Phi}|\alpha)= p(\mathbf{x}_c|\bm{\Phi},\alpha)p_o(\bm{\Phi})$.
\section{Performance bounds}
\label{sec:perf_bounds}
We derive BCRLB for each of these three modes based on prior information on $\bm{\Phi}$. In the performance bound derivation, we focus on the case when both radar and target are static which simplifies the analysis yet still provides the insights into the performance of different systems. 
\begin{theorem}[Point-cloud fusion BCRLB]
Given the prior $\bm{\Phi} \sim N(\bm{\Phi}_o,\mathbf{R_o})$, the BCRLB for point-cloud fusion is the inverse of the Fisher Information Matrix (FIM)\par\noindent\small
\begin{align}
    \mathbf{F}_{\Phi} = \mathbb{E}\{\mathbf{F}_l\}+\mathbf{F}_0,
\end{align}\normalsize
where $\mathbf{F}_l$ and $\mathbf{F}_0$ are determined by deterministic CRLB and prior information, respectively.
\end{theorem}
\begin{IEEEproof}
For conventional processing, the deterministic CRLB is based on the likelihood function $p(\mathbf{x}|\bm{\Phi})$ as\par\noindent\small
\begin{equation}
    \mathbf{C}_{\hat{\bm{\Phi}}}=\mathbb{E}_{\mathbf{x}|\bm{\Phi}}\left\{(\bm{\Phi}-\hat{\bm{\Phi}})(\bm{\Phi}-\hat{\bm{\Phi}})^T\right\}\succeq \mathbf{F}_l^{-1}, 
\end{equation}\normalsize
where the FIM $\mathbf{F}_l$ is\par\noindent\small
\begin{align}
    \mathbf{F}_l&=-\mathbb{E}\left\{\frac{\partial^2 \ln p(\mathbf{x}|\bm{\Phi})}{\partial \bm{\Phi}^2}\right\}\nonumber\\
    &=\mathbb{E}\left\{\left(\frac{\partial \ln p(\mathbf{x}|\bm{\Phi})}{\partial \bm{\Phi}}\right)\left(\frac{\partial \ln p(\mathbf{x}|\bm{\Phi})}{\partial \bm{\Phi}}\right)^T\right\}.
\end{align}\normalsize
The BCRLB is based on posterior distribution $p(\bm{\Phi}|\mathbf{x})=p(\mathbf{x}|\bm{\Phi})p_o(\bm{\Phi})$,\par\noindent\small
\begin{equation}
    \mathbf{C}_{\hat{\bm{\Phi}}}=\mathbb{E}_{\mathbf{x},\bm{\Phi}}\left\{(\bm{\Phi}-\hat{\bm{\Phi}})(\bm{\Phi}-\hat{\bm{\Phi}})^T\right\}\succeq \mathbf{F}_{\Phi}^{-1}, 
\end{equation}\normalsize
where the Bayesian FIM $\mathbf{F}$ for point-cloud fusion processing is \cite{tichavsky1998posterior}\par\noindent\small 
\begin{align}
    \mathbf{F}_{\Phi}&=-\mathbb{E}_{\mathbf{x},\bm{\Phi}}\left\{\frac{\partial^2 \ln p(\bm{\Phi},\mathbf{x})}{\partial \bm{\Phi}^2}\right\} \nonumber\\
    &=-\mathbb{E}_{\mathbf{x},\bm{\Phi}}\left\{\frac{\partial^2 \ln p(\mathbf{x}|\bm{\Phi})}{\partial \bm{\Phi}^2}\right\}-\mathbb{E}_{\bm{\Phi}}\left\{\frac{\partial ^2 \ln p_o(\bm{\Phi})}{\partial \bm{\Phi}^2}\right\} \nonumber\\
    &=\mathbb{E}_{\bm{\Phi}}\left\{-\mathbb{E}_{\mathbf{x}|\bm{\Phi}}\left\{\frac{\partial^2 \ln p(\mathbf{x}|\bm{\Phi})}{\partial \bm{\Phi}^2}\right\}\right\}-\mathbb{E}_{\bm{\Phi}}\left\{\frac{\partial ^2 \ln p_o(\bm{\Phi})}{\partial \bm{\Phi}^2}\right\} \nonumber\\
    &=\mathbb{E}_{\bm{\Phi}}\{\mathbf{F}_l\} + \mathbf{F}_o.
\end{align}\normalsize
Using the noise covariance, $\mathbf{F}_l=\mathbf{g}^T\mathbf{R}^{-1}_n\mathbf{g}$, with $\mathbf{g}=\frac{\partial \mathbf{f}(\bm{\Phi})}{\partial \bm{\Phi}}$ where
\begin{equation}
    \frac{\partial r_q}{\partial \bm{\Phi}}=\left[\frac{x-x_q}{r_q},  \frac{y-y_q}{r_q},\frac{z-z_q}{r-q} \right],
\end{equation}
\par\noindent
\begin{align}
    \frac{\partial \theta _q}{\partial \bm{\Phi}}&=\left[\frac{\partial \theta _q}{\partial x},\frac{\partial \theta _q}{\partial y},\frac{\partial \theta _q}{\partial z}\right]. 
\end{align}
with \par\noindent
\begin{align}
    \frac{\partial \theta _q}{\partial x}&=\left(1+\left(\frac{y-y_q}{x-x_q}\right)^2\right)^{-1}\frac{y_q-y}{(x-x_q)^2}, \\
    \frac{\partial \theta _q}{\partial y}&=\left(1+\left(\frac{y-y_q}{x-x_q}\right)^2\right)^{-1}\frac{1}{x-x_q},\\
    \frac{\partial \theta _q}{\partial z}&=0.
\end{align}
and \par\noindent
\begin{align}
    \frac{\partial \phi _q}{\partial \bm{\Phi}}&=\left[\frac{\partial \phi _q}{\partial x},\frac{\partial \phi _q}{\partial y},\frac{\partial \phi _q}{\partial z}\right]. 
\end{align}
with\noindent
\begin{align}
\frac{\partial \phi _q}{\partial x}&=\left(1+\left(\frac{z-z_q}{((x-x_{q})^2+(y-y_{q})^2)^{1/2}}\right)^2\right)^{-1}\nonumber\\
    &\frac{(x_q-x)(z-z_q)}{((x-x_{q})^2+(y-y_{q})^2)^{3/2}},\\
\frac{\partial \phi _q}{\partial y}&=\left(1+\left(\frac{z-z_q}{((x-x_{q})^2+(y-y_{q})^2)^{1/2}}\right)^2\right)^{-1}\nonumber\\
&\frac{(y_q-y)(z-z_q)}{((x-x_{q})^2+(y-y_{q})^2)^{3/2}},\\
\frac{\partial \theta _q}{\partial z}&=\left(1+\left(\frac{z-z_q}{((x-x_{q})^2+(y-y_{q})^2)^{1/2}}\right)^2\right)^{-1}\nonumber\\
&\frac{1}{((x-x_{q})^2+(y-y_{q})^2)^{1/2}}.
\end{align}
Using the prior $\bm{\Phi} \sim N(\bm{\Phi}_o,\mathbf{R_o})$ yields $F_o=R_o^{-1}$.
\end{IEEEproof}

\begin{theorem}[Non-coherent processing BCRLB]
Given the deterministic but unknown nuisance parameters $\bm{\alpha}$, the BCRLB of $\bm{\Phi}$ in case of non-coherent processing is conditioned on $\bm{\alpha}$ as\par\noindent\small
\begin{equation}
    \mathbf{F}_{\bm{\Phi}}^{-1}=(\mathbb{E}_{\bm{\Phi}}\left\{\mathbf{F}_{\bm{\Phi},\bm{\Phi}}\right\}+\mathbf{F}_o-\mathbb{E}_{\bm{\Phi}}\left\{\mathbf{F}_{\bm{\Phi},\bm{\alpha}}\right\}\mathbf{F}^{-1}_{\bm{\alpha},\bm{\alpha}}\mathbb{E}_{\bm{\Phi}}\left\{\mathbf{F}_{\bm{\alpha},\bm{\Phi}}\right\})^{-1}.
    \label{eq:bcrlbNCP}
\end{equation}\normalsize
\end{theorem}

\begin{IEEEproof}
The BCRLB conditioned on $\bm{\alpha}$ is\par\noindent\small
\begin{equation}
    \mathbf{C}_{\bm{\Phi}|\bm{\alpha}}=\mathbb{E}_{\mathbf{x}_c,\bm{\Phi}|\bm{\alpha}}\{(\bm{\Phi}-\hat{\bm{\Phi}})(\bm{\Phi}-\hat{\bm{\Phi}})^T\}\succeq \mathbf{F}_{\bm{\Phi}}^{-1}.
\end{equation}\normalsize
The nuisance parameters make it difficult to directly arrive at the BCRLB of $\bm{\Phi}$. Therefore, we derive it from the BCRLB of $\bm{\Phi}_1$, which is a hybrid lower bound because of presence of both random and deterministic parameters. The BCRLB of $\bm{\Phi}_1$ is\par\noindent\small
\begin{equation}
    \mathbf{C}_{\bm{\Phi}_1}=\mathbb{E}_{\mathbf{x}_c,\bm{\Phi}|\bm{\alpha}}\{(\bm{\Phi}_1-\hat{\bm{\Phi}}_1)(\bm{\Phi}_1-\hat{\bm{\Phi}}_1)^T\}\succeq \mathbf{F}^{-1},
\end{equation}\normalsize
where $\mathbf{F}\triangleq \mathbb{E}_{\bm{\Phi}}\{\mathbf{F}_l\}+\mathbf{F}_p$ with \par\noindent\small
\begin{align}
    \mathbf{F}_{l}&=-\mathbb{E}_{\mathbf{x}_{nc}|\bm{\Phi};\bm{\alpha}}\left\{\frac{\partial^2 \ln p(\mathbf{x}_{nc}|\bm{\Phi};\bm{\alpha})}{\partial \bm{\Phi}_1^2}\right\},\label{eq:FlmatNCP}\\
    \mathbf{F}_p&=-\mathbb{E}_{\bm{\Phi}_1}\left\{\frac{\partial^2\ln p_o(\bm{\Phi}_1)}{\partial \bm{\Phi}_1^2}\right\}=\begin{bmatrix}
         \mathbf{F}_o & \mathbf{0} \\
         \mathbf{0} & \mathbf{0}
    \end{bmatrix}.
\end{align}\normalsize
In block matrix form,\par\noindent\small
\begin{equation}
    \mathbf{F}_l=\begin{bmatrix}
         \mathbf{F}_{\bm{\Phi},\bm{\Phi}} & \mathbf{F}_{\bm{\Phi},\bm{\alpha}} \\
         \mathbf{F}_{\bm{\alpha},\bm{\Phi}} & \mathbf{F}_{\bm{\alpha},\bm{\alpha}}
    \end{bmatrix},
\end{equation}\normalsize
so that\par\noindent\small
\begin{equation}
    \mathbf{F}=\begin{bmatrix}
         \mathbb{E}_{\bm{\Phi}}\left\{\mathbf{F}_{\bm{\Phi},\bm{\Phi}}\right\}+\mathbf{F}_o & \mathbb{E}_{\bm{\Phi}}\left\{\mathbf{F}_{\bm{\Phi},\bm{\alpha}}\right\} \\
         \mathbb{E}_{\bm{\Phi}}\left\{\mathbf{F}_{\bm{\alpha},\bm{\Phi}}\right\}& \mathbf{F}_{\bm{\alpha},\bm{\alpha}}
    \end{bmatrix}.
    \label{eq:Fmat}
\end{equation}\normalsize
Taking the Shur complement of (\ref{eq:Fmat}) completes the proof.
\end{IEEEproof}
In order to evaluate (\ref{eq:bcrlbNCP}), we need to explicitly derive (\ref{eq:FlmatNCP}). From (\ref{eq:RxSigModNCP}), we have $
    \mathbf{u}_x\triangleq \frac{\partial \mathbf{h}(\bm{\Phi})}{\partial x}=\mathbf{h}(\bm{\Phi})\circ \mathbf{G}_x(\bm{\Phi})$, with $\mathbf{G}_x(\bm{\Phi})=[G^x_{1,1,1,1},\cdots,G^x_{q,m,n,n_s},\cdots,G^x_{Q,M,N,N_s}]$ such that\par\noindent\small
\begin{align}
    G^x_{q,m,n,n_s}&=j2\pi\left(\frac{f_c^q}{c}+\frac{B_rt_{n_s}}{c^2}\right)\frac{\partial g_{qmn}}{\partial x},\\
    \frac{\partial g_{qmn}}{\partial x}& =\frac{x-x_{qn}}{((x_{qn}-x)^2+(y_{q,n}-y)^2+(z_{q,n}-z)^2)^{1/2}}\nonumber\\
    &+\frac{x-x_{qm}}{((x_{qm}-x)^2+(y_{q,m}-y)^2+(z_{q,n}-z)^2)^{1/2}}.
\end{align}\normalsize
where $g_{qmn}$ is given by
\begin{equation}
\begin{split}
    g_{qmn}&=((x_{qn}-x)^2+(y_{q,n}-y)^2+(z_{q,n}-z)^2)^{1/2} \\
    &+((x_{q,m}-x)^2+(y_{q,m}-y)^2+(z_{q,m}-z)^2)^{1/2} \\
\end{split}
\end{equation}
The $\mathbf{u}_y\triangleq \frac{\partial \mathbf{h}(\bm{\Phi})}{\partial y}=\mathbf{h}(\bm{\Phi})\circ \mathbf{G}_y(\bm{\Phi})$ is defined similarly.

\begin{equation}
    G^y_{qmnn_s}=j2\pi\left(\frac{f_c}{c}+\frac{B_rt_{n_s}}{c^2}\right)\frac{\partial g_{qmn}}{\partial y}
\end{equation}

\begin{equation}
\begin{split}
    \frac{\partial g_{qmn}}{\partial y}&=\frac{y-y_{q,n}}{((x_{q,n}-x)^2+(y_{q,n}-y)^2+(z_{q,n}-z)^2)^{1/2}}\\ &+\frac{y-y_{q,m}}{((x_{q,m}-x)^2+(y_{q,m}-y)^2+(z_{q,m}-z))^{1/2}}\\
\end{split}
\end{equation}
The $\mathbf{u}_z\triangleq \frac{\partial \mathbf{h}(\bm{\Phi})}{\partial z}=\mathbf{h}(\bm{\Phi})\circ \mathbf{G}_z(\bm{\Phi})$ is defined similarly.

\begin{equation}
    G^z_{qmnn_s}=j2\pi\left(\frac{f_c}{c}+\frac{B_rt_{n_s}}{c^2}\right)\frac{\partial g_{qmn}}{\partial z}
\end{equation}
\begin{equation}
\begin{split}
    \frac{\partial g_{qmn}}{\partial z}&=\frac{z-z_{q,n}}{((x_{q,n}-x)^2+(y_{q,n}-y)^2+(z_{q,n}-z)^2)^{1/2}}\\ &+\frac{z-z_{q,m}}{((x_{q,m}-x)^2+(y_{q,m}-y)^2+(z_{q,m}-z))^{1/2}}\\
\end{split}
\end{equation}

Let $\mathbf{A}=\mathrm{diag}\{\bm{\alpha}\otimes \mathbf{1}_{MNN_s}\}$ and $\mathbf{R}_A=E[\mathbf{A}\mathbf{R}_{nc}^{-1}\mathbf{A}]$. Assuming that the noise components over different radar sensors are i.i.d., the inverse of the noise covariance matrix is $\mathbf{R}_{nc}^{-1}=\mathrm{diag}\{\mathbf{R_{1,1}^{-1}},\cdots,\mathbf{R}_{Q,Q}^{-1}\}$, where $\mathbf{R}_{q,q}^{-1}$ is the inverse of the noise covariance matrix of the $q$-th radar. We define $h_q(\bm{\Phi})\triangleq {\mathcal{S}_q}(\bm{\Phi})$, meaning that $h_q(\bm{\Phi})$ consists of the rows as indexed in the set of $\mathcal{S}_q$ of $\mathbf{h}(\bm{\Phi})$, with $\mathcal{S}_q=\{(q-1)MNN_s+1,\cdots,qMNN_s\}$. In other words, $\mathbf{h}_q(\bm{\Phi})=[h_{q,1,1}(t_1),\cdots,h_{q,M,N}(t_{N_s})]^T$. Similarly, define $\mathbf{u}_x^q=\mathbf{u}_x(\mathcal{S}_q,1)$ and $\mathbf{u}_y^q=\mathbf{u}_y(\mathcal{S}_q,1)$. Then, the FIM is 
\begin{widetext}
\begin{align}
    \mathbf{F}_l &= \begin{bmatrix}
    \mathbf{F}_l(x,x) & \mathbf{F}_l(x,y)& \mathbf{F}_l(x,z) & \mathbf{F}_l(x,\alpha_r^1) & \cdots & \mathbf{F}_l(x,\alpha_r^Q) & \mathbf{F}_l(x,\alpha_i^1) & \cdots & \mathbf{F}_l(x,\alpha_i^Q)\\
    \mathbf{F}_l(y,x) & \mathbf{F}_l(y,y) & \mathbf{F}_l(y,z)& \mathbf{F}_l(y,\alpha_r^1) & \cdots & \mathbf{F}_l(y,\alpha_r^Q) & \mathbf{F}_l(y,\alpha_i^1) & \cdots & \mathbf{F}_l(y,\alpha_i^Q)\\
    \mathbf{F}_l(z,x) & \mathbf{F}_l(z,y) & \mathbf{F}_l(z,z)& \mathbf{F}_l(z,\alpha_r^1) & \cdots & \mathbf{F}_l(z,\alpha_r^Q) & \mathbf{F}_l(z,\alpha_i^1) & \cdots & \mathbf{F}_l(z,\alpha_i^Q)\\
    \mathbf{F}_l(\alpha_r^1,x) & \mathbf{F}_l(\alpha_r^1,y)& \mathbf{F}_l(\alpha_r^1,z) & \mathbf{F}_l(\alpha_r^1,\alpha_r^1) & \cdots & \mathbf{F}_l(\alpha_r^1,\alpha_r^Q) & \mathbf{F}_l(\alpha_r^1,\alpha_i^1) & \cdots & \mathbf{F}_l(\alpha_r^1,\alpha_i^Q)\\
    \vdots & \vdots & \vdots & \vdots & \vdots & \vdots & \vdots & \vdots & \vdots\\
    \mathbf{F}_l(\alpha_r^Q,x) & \mathbf{F}_l(\alpha_r^Q,y)& \mathbf{F}_l(\alpha_r^Q,z) & \mathbf{F}_l(\alpha_r^Q,\alpha_r^1) & \cdots & \mathbf{F}_l(\alpha_r^Q,\alpha_r^Q) & \mathbf{F}_l(\alpha_r^Q,\alpha_i^1) & \cdots & \mathbf{F}_l(\alpha_r^Q,\alpha_i^Q)\\
    \mathbf{F}_l(\alpha_i^1,x) & \mathbf{F}_l(\alpha_i^1,y)& \mathbf{F}_l(\alpha_i^1,z) & \mathbf{F}_l(\alpha_i^1,\alpha_r^1) & \cdots &\mathbf{F}_l(\alpha_i^1,\alpha_r^Q) &
    \mathbf{F}_l(\alpha_i^1,\alpha_i^1) & \cdots &\mathbf{F}_l(\alpha_i^1,\alpha_i^Q)\\
    \vdots & \vdots & \vdots & \vdots & \vdots & \vdots & \vdots & \vdots & \vdots\\
    \mathbf{F}_l(\alpha_i^Q,x) & \mathbf{F}_l(\alpha_i^Q,y)& \mathbf{F}_l(\alpha_i^Q,z) & \mathbf{F}_l(\alpha_i^Q,\alpha_r^1) & \cdots &\mathbf{F}_l(\alpha_i^Q,\alpha_r^Q) &
    \mathbf{F}_l(\alpha_i^Q,\alpha_i^1) & \cdots &\mathbf{F}_l(\alpha_i^Q,\alpha_i^Q)
    \end{bmatrix}
\end{align}
\end{widetext}
with  
\begin{align}
\mathbf{F}_l(x,x)&=2\mathbf{u}_x^H\mathbf{R}_A\mathbf{u}_x,\nonumber\\
\mathbf{F}_l(x,y)&=2\Re\{\mathbf{u}_x^H\mathbf{R}_A\mathbf{u}_y\},\nonumber\\
\mathbf{F}_l(x,z)&=2\Re\{\mathbf{u}_x^H\mathbf{R}_A\mathbf{u}_z\},\nonumber\\
\mathbf{F}_l(y,y)&=2\mathbf{u}_y^H\mathbf{R}_A\mathbf{u}_y, \nonumber\\
\mathbf{F}_l(y,z)&=2\mathbf{u}_y^H\mathbf{R}_A\mathbf{u}_z, \nonumber\\
\mathbf{F}_l(x,\alpha_r^q)&=2\Re\{(\alpha_q\mathbf{u}_x^q)^H\mathbf{R}_{q,q}^{-1}\mathbf{h}_q(\bm{\Phi})\},\nonumber\\
\mathbf{F}_l(y,\alpha_r^q)&=2\Re\{(\alpha_q\mathbf{u}_y^q)^H\mathbf{R}_{q,q}^{-1}\mathbf{h}_q(\bm{\Phi})\},\nonumber\\
\mathbf{F}_l(z,\alpha_r^q)&=2\Re\{(\alpha_q\mathbf{u}_z^q)^H\mathbf{R}_{q,q}^{-1}\mathbf{h}_q(\bm{\Phi})\},\nonumber\\
\mathbf{F}_l(x,\alpha_i^q)&=2\Im\{\mathbf{h}_q(\bm{\Phi})^H\mathbf{R}_{q,q}^{-1}(\alpha_q\mathbf{u}_x)\},\nonumber\\
\mathbf{F}_l(y,\alpha_i^q)&=2\Im\{\mathbf{h}_q(\bm{\Phi})^H
\mathbf{R}_{q,q}^{-1}(\alpha_q\mathbf{u}_y^q)\},\nonumber\\
\mathbf{F}_l(z,\alpha_i^q)&=2\Im\{\mathbf{h}_q(\bm{\Phi})^H
\mathbf{R}_{q,q}^{-1}(\alpha_q\mathbf{u}_z^q)\},\nonumber\\
\mathbf{F}_l(\alpha_r^q,\alpha_r^q)&=2\mathbf{h}_q(\bm{\Phi})^H\mathbf{R}_{q,q}^{-1}\mathbf{h}_q(\bm{\Phi}),\nonumber\\
\mathbf{F}_l(\alpha_i^q,\alpha_i^q)&=2\mathbf{h}_q(\bm{\Phi})^H\mathbf{R}_{q,q}^{-1}\mathbf{h}_q(\bm{\Phi}),\nonumber\\
\mathbf{F}_l(\alpha_r^q,\alpha_i^q)&=\mathbf{F}_l(\alpha_r^{q_1},\alpha_r^{q_2})=\mathbf{F}_l(\alpha_i^{q_1},
\alpha_i^{q_2})=\mathbf{F}_l(\alpha_r^{q_1},\alpha_i^{q_2})=0.
\end{align}

\begin{theorem}[Coherent processing BCRLB]
Given the deterministic but unknown nuisance parameter $\alpha$, the BCRLB of $\bm{\Phi}$ in case of coherent processing is conditioned on $\alpha$ as\par\noindent\small
\begin{equation}
    \mathbf{F}_{\bm{\Phi}}^{-1}=(\mathbb{E}_{\bm{\Phi}}\left\{\mathbf{F}_{\bm{\Phi},\bm{\Phi}}\right\}+\mathbf{F}_o-\mathbb{E}_{\bm{\Phi}}\left\{\mathbf{F}_{\bm{\Phi},\alpha}\right\}\mathbf{F}^{-1}_{\alpha,\alpha}\mathbb{E}_{\bm{\Phi}}\left\{\mathbf{F}_{\alpha,\bm{\Phi}}\right\})^{-1}.
    \label{eq:bcrlbCP}
\end{equation}\normalsize
\end{theorem}

\begin{IEEEproof}
The proof follows by substituting $\alpha_q=\alpha, q=1,\cdots,Q$ in the BCRLB of non-coherent processing. 
Define $\bm{\Phi}_2=[\bm{\Phi},\alpha]$, then following the development in the non-coherent processing, the BCRLB is given by
\begin{equation}
    C_{\bm{\Phi}_2}=\mathbb{E}_{\mathbf{x}_c,\bm{\Phi}|\alpha}[(\bm{\Phi}_2-\hat{\bm{\Phi}}_2)(\bm{\Phi}_2-\hat{\bm{\Phi}}_2)^T]\succeq F^{-1},
\end{equation}
where $F\triangleq \mathbb{E}_{\bm{\Phi}}[F_l]+F_p$ with $F_l$ given by
\begin{equation}
    F_{l}=-\mathbb{E}_{\mathbf{x}_{nc}|\bm{\Phi};\alpha}\left\{\frac{\partial^2 \textit{ln} p(\mathbf{x}_{nc}|\bm{\Phi};\alpha)}{\partial \bm{\Phi}_2^2}\right\},
    \label{eq:FlmatCP}
\end{equation}
and $F_p$ given by
\begin{equation}
    F_p=-\mathbb{E}_{\bm{\Phi}_2}\left\{\frac{\partial^2\ln p_o(\bm{\Phi}_2)}{\partial \bm{\Phi}_2^2}\right\}=\begin{bmatrix}
         F_o & \mathbf{0} \\
         \mathbf{0} & \mathbf{0}
    \end{bmatrix}.
\end{equation}
Similarly, write the likelihood Fisher information into a block matrix form,
\begin{equation}
    F_l=\begin{bmatrix}
         F_{\bm{\Phi},\bm{\Phi}} & F_{\bm{\Phi},\alpha} \\
         F_{\alpha,\bm{\Phi}} & F_{\alpha,\alpha}
    \end{bmatrix},
\end{equation}
then,
\begin{equation}
    F=\begin{bmatrix}
         \mathbb{E}_{\bm{\Phi}}\left\{F_{\bm{\Phi},\bm{\Phi}}\right\}+F_o & \mathbb{E}_{\bm{\Phi}}\left\{F_{\bm{\Phi},\alpha}\right\} \\
         \mathbb{E}_{\bm{\Phi}}\left\{F_{\alpha,\bm{\Phi}}\right\}& F_{\alpha,\alpha}
    \end{bmatrix}.
    \label{eq:Fmatcp}
\end{equation}
Based on (\ref{eq:Fmatcp}), applying the Schur complement theory, we can write the BCRLB conditioned on $\alpha$  as:
\begin{equation}
    F_{\bm{\Phi}}^{-1}=(\mathbb{E}_{\bm{\Phi}}\left\{F_{\bm{\Phi},\bm{\Phi}}\right\}+F_o-\mathbb{E}_{\bm{\Phi}}\left\{F_{\bm{\Phi},\alpha}\right\}F^{-1}_{\alpha,\alpha}\mathbb{E}_{\bm{\Phi}}\left\{F_{\alpha,\bm{\Phi}}\right\})^{-1}.
    \label{eq:bcrlbCP1}
\end{equation}

To complete the BCRLB derivation, the likelihood Fisher information matrix is
\begin{widetext}
\begin{align}
    \mathbf{F}_l &= \begin{bmatrix}
    \mathbf{F}_l(x,x) & \mathbf{F}_l(x,y) & \mathbf{F}_l(x,z)& \mathbf{F}_l(x,\alpha_r) &  \mathbf{F}_l(x,\alpha_i)\\
    \mathbf{F}_l(y,x) & \mathbf{F}_l(y,y)& \mathbf{F}_l(y,z) & \mathbf{F}_l(y,\alpha_r) &  \mathbf{F}_l(y,\alpha_i)\\
    \mathbf{F}_l(z,x) & \mathbf{F}_l(z,y)& \mathbf{F}_l(z,z) & \mathbf{F}_l(z,\alpha_r) &  \mathbf{F}_l(z,\alpha_i)\\
    \mathbf{F}_l(\alpha_r,x) & \mathbf{F}_l(\alpha_r,y) &\mathbf{F}_l(\alpha_r,z) & \mathbf{F}_l(\alpha_r,\alpha_r) &  \mathbf{F}_l(\alpha_r,\alpha_i)\\
    \mathbf{F}_l(\alpha_i,x) & \mathbf{F}_l(\alpha_i,y) &\mathbf{F}_l(\alpha_i,z) & \mathbf{F}_l(\alpha_i,\alpha_r) &  \mathbf{F}_l(\alpha_i,\alpha_i)
    \end{bmatrix}\nonumber\\
    &= \begin{bmatrix}
    2\sigma^2\mathbf{u}_x^H\mathbf{R}_c^{-1}\mathbf{u}_x & 2\sigma^2\Re\{\mathbf{u}_x^H\mathbf{R}_c^{-1}\mathbf{u}_y\}&
    2\sigma^2\Re\{\mathbf{u}_x^H\mathbf{R}_c^{-1}\mathbf{u}_z\}& 2\Re \{\alpha^H\mathbf{u}^H_x\mathbf{R}_c^{-1}\mathbf{h}(\bm{\Phi})\} & 2\Im \{\mathbf{h}(\bm{\Phi})^H\mathbf{R}_c^{-1}\mathbf{u}_x\alpha\}\\
    2\sigma^2\Re\{\mathbf{u}_x^H\mathbf{R}_c^{-1}\mathbf{u}_y\} & 2\sigma^2\mathbf{u}_y^H\mathbf{R}_c^{-1}\mathbf{u}_y &2\sigma^2\Re\{\mathbf{u}_y^H\mathbf{R}_c^{-1}\mathbf{u}_z\}&2\Re \{\alpha^H\mathbf{u}^H_y\mathbf{R}_c^{-1}h(\bm{\Phi})\} & 2\Im \{\mathbf{h}(\bm{\Phi})^H\mathbf{R}_c^{-1}\mathbf{u}_y\alpha\}\\
    2\sigma^2\Re\{\mathbf{u}_x^H\mathbf{R}_c^{-1}\mathbf{u}_z\}&
    2\sigma^2\Re\{\mathbf{u}_y^H\mathbf{R}_c^{-1}\mathbf{u}_z\}&
    2\sigma^2\mathbf{u}_z^H\mathbf{R}_c^{-1}\mathbf{u}_z&
    2\Re \{\alpha^H\mathbf{u}^H_z\mathbf{R}_c^{-1}h(\bm{\Phi})\} & 2\Im \{\mathbf{h}(\bm{\Phi})^H\mathbf{R}_c^{-1}\mathbf{u}_z\alpha\}\\
   2\Re \{\alpha^H\mathbf{u}^H_x\mathbf{R}_c^{-1}\mathbf{h}(\bm{\Phi})\} & 2\Re \{\alpha^H\mathbf{u}^H_y\mathbf{R}_c^{-1}h(\bm{\Phi})\} & 2\Re \{\alpha^H\mathbf{u}^H_z\mathbf{R}_c^{-1}h(\bm{\Phi})\} & 2\mathbf{h}(\bm{\Phi})^H\mathbf{R}_c^{-1}\mathbf{h}(\bm{\Phi}) &  0\\
    2\Im \{\mathbf{h}(\bm{\Phi})^H\mathbf{R}_c^{-1}\mathbf{u}_x\alpha\} & 2\Im \{\mathbf{h}(\bm{\Phi})^H\mathbf{R}_c^{-1}\mathbf{u}_y\alpha\} & 2\Im \{\mathbf{h}(\bm{\Phi})^H\mathbf{R}_c^{-1}\mathbf{u}_z\alpha\}& 0 &  2\mathbf{h}(\bm{\Phi})^H\mathbf{R}_c^{-1}\mathbf{h}(\bm{\Phi})
    \end{bmatrix}
\end{align}
\end{widetext}
\end{IEEEproof}
\section{Displaced Sensor Imaging}
\label{sec:displ}
It follows from the performance bounds (see also Section~\ref{subsec:num_exp_perf}) that the point-cloud fusion method has the least computational load as well as the worst performance when compared with both non-coherent processing and coherent processing. Based on asynchronous clock assumption, coherent processing is not directly usable. Therefore, we first focus on developing displaced sensor imaging for the non-coherent mode using reduced-rate sensing and block-sparse reconstruction. 
Then, we devise a more effective coherent imaging process taking into account the synchronization errors. 
\subsection{Non-coherent imaging}
Beginning with the basic signal model in (\ref{eq:sqmntkv1}), define\par\noindent\small
\begin{equation}
   \begin{split}
    h_{qmnkn_s}&=e^{-\mathrm{j}2\pi f_c^qg_{q,m,n}/c}\\
    &e^{-\mathrm{j}2\pi\left(\frac{2f_c^qv}{c}+\frac{B_rg_{qmn}}{c}\right)n_sT_s}e^{-\mathrm{j}2\pi\frac{2f_c^qvT_r}{c}(n-1+(k-1)N))}.
    \end{split}
\end{equation}\normalsize
The received echo from the target is $y_{qmn}(n_s,k)=\alpha_qh_{qmnkn_s}$. Stacking all indices gives $\mathbf{y}_{nc}=[\alpha_1h_{11111,\cdots,\alpha_Qh_{QMNKN_s}}]^T\triangleq (\mathbf{\alpha}\otimes\mathbf{I}_{MNKN_s\times 1})\circ\mathbf{h}(\bm{\Phi})$, where $\mathbf{h}(\bm{\Phi})=[h_{11111},\cdots,h_{QMNKN_s}]^T$. In the presence of noise, the received signal from single target becomes\par\noindent\small
\begin{equation}
\begin{split}
\mathbf{r}_{nc}&=(\bm{\alpha}\otimes I_{MNKN_s\times 1})\circ\mathbf{h}(\bm{\Phi})+\mathbf{n}_{nc}
\triangleq \mathbf{h}_{\alpha}(\Phi)+\mathbf{n}_c.
\end{split}
\end{equation}\normalsize
Assume the field of view is divided into multiple sets of $\Phi_l=(x_l,y_l,z_l)^T, l=1,\cdots,L$. Defines $\mathbf{h}_{\alpha}(\bm{\Phi}_l)=(\bm{\alpha}_l\otimes I_{MNKN_s\times 1})\circ\mathbf{h}(\bm{\Phi}_l)$. Then, for multiple targets, the received signal is\par\noindent\small
\begin{equation}
\mathbf{r}_{nc}=\mathbf{H}_{\alpha}\mathbf{c}+\mathbf{n}_c,
\label{eq:nonSPH}
\end{equation}\normalsize
where $\mathbf{H}_{\alpha}=[\mathbf{h}_{\alpha}(\bm{\Phi}_1),\cdots,\mathbf{h}_{\alpha}(\bm{\Phi}_L)]$ and $\mathbf{c}=[c_1, \cdots, c_L]^T$ has elements $1$ or $0$ depending on the presence or absence of a target. Here, the reflection coefficient is subsumed into $\mathbf{H}_{\alpha}$, making it difficult to directly estimate the coefficient. In the sequel, we present a model in which the coefficient is easily estimated. 

From (16), define $\mathbf{A}=\mathrm{diag}\{\bm{\alpha}\}$ so that the received signal for single target becomes\par\noindent\small
\begin{equation}
\begin{split}
\mathbf{r}_{nc}&=(\bm{\alpha}\otimes I_{MNKN_s\times 1})\circ\mathbf{h}(\bm{\Phi})+\mathbf{n}_{nc}\\
&=\mathrm{vec}\{[\mathbf{h}_1(\Phi), \cdots, \mathbf{h}_Q(\Phi)]\mathbf{A}\}+\mathbf{n}_{nc}\\
&=(\mathbf{I}_Q\otimes \mathbf{H})\mathrm{vec}\{\mathbf{A}\}+\mathbf{n}_{nc}=(\mathbf{I}_Q\otimes \mathbf{H})\mathbf{S}\mathbf{\alpha}+\mathbf{n}_{nc}\\
&\triangleq \mathbf{H}_s(\Phi)\mathbf{\alpha}+\mathbf{n}_{nc},
\end{split}
\label{eq:nonSPa}
\end{equation}\normalsize
where $\mathbf{S}$ is the selection matrix such that $\mathrm{vec}\{\mathbf{A}\}=\mathbf{S}\mathbf{\alpha}$. Define the point target response function (PTRF) of non-coherent processing as\par\noindent\small
\begin{equation}
\begin{split}
F(\Phi)&=(\mathbf{H}_s(\Phi)\mathbf{\alpha})^H(\mathbf{H}_s(\Phi)\mathbf{\alpha})\\
&=\sum_{q=1}^Q|\alpha_q|^2\mathbf{h}_q^H(\Phi)\mathbf{h}_q(\Phi).
\end{split}
\label{eq:NCP-PTRF}
\end{equation}\normalsize
Note that, although there is no processing loss in signal integration, the resolution is largely limited by that of each individual sensor. Later, in Section~\ref{subsec:ptrf}, we compare this with the PTRF of coherent processing.

Generalizing the above for multiple targets yields\par\noindent\small
\begin{equation}
\begin{split}
\mathbf{r}_{nc}&=\sum_{p=1}^{P}\mathbf{H}_s(\Phi_{t_p})\mathbf{\alpha}_{p}=\tilde{\mathbf{H}}_s\tilde{\bm{\alpha}}+\mathbf{n}_{nc},
\end{split}
\label{eq:nonSPaP}
\end{equation}\normalsize
where $\tilde{\mathbf{H}}_s=[\mathbf{H}_s(\Phi_{t_1}),\cdots,\mathbf{H}_s(\Phi_{t_P})]$ and $\tilde{\bm{\alpha}}=[\bm{\alpha}_{1};\cdots,\bm{\alpha}_{P}]$ with $\bm{\alpha}_{p}=[\alpha_{{p},1};\cdots;\alpha_{{p},Q}]$ standing for the $Q$ signal reflection coefficient for the $p-$th target. It follows from (\ref{eq:nonSPaP}) that, similar to conventional imaging, the target position is captured in $\Phi_{t_p}$ but, differently, the reflectivity is captured in a vector of same size as the number of displaced sensors. 

From (\ref{eq:nonSPaP}), discretizing the field of view into $\Phi_l$ results in the following sparse localization model\par\noindent\small
\begin{equation}
\begin{split}
\mathbf{r}_{nc}&=\sum_{p=1}^{P}\mathbf{H}_s(\Phi_p)\mathbf{\alpha}_p=\mathbf{H}\mathbf{b}+\mathbf{n}_{nc},
\end{split}
\label{Eq:sigModCS}
\end{equation}\normalsize
where $\mathbf{H}=[\mathbf{H}_s(\Phi_{1}),\cdots,\mathbf{H}_s(\Phi_{L})]$ is the basis matrix and $\mathbf{b}=[\mathbf{b}_1;\cdots;\mathbf{b}_L]$ is a block-wise sparse vector, with $\mathbf{b}_l=\mathbf{0}$ if there is no target at $\Phi_l$ and $\mathbf{b}_l=\mathbf{\alpha}_p$ only if the $p-$th target is located at $\Phi_l$. 
To recover $\mathbf{b}$ in (\ref{Eq:sigModCS}), we adopt the following block-sparse image recovery algorithms.
\subsubsection{$\ell_1$-norm optimization for direct sparse-recovery}
The common $\ell_1$-norm minimization \cite{foucart2013invitation} provides the estimate $\hat{\mathbf{b}}$ by solving the optimization problem
\begin{equation}
\underset{\mathbf{b}}{\text{minimize}} \quad ||\mathbf{b}||_1 \quad \text{subject to} \quad ||\mathbf{H}\mathbf{b}-\mathbf{r}_{nc}||_2\leqslant \epsilon,
\end{equation}
where $\epsilon$ is some constant related to the noise variance. 
For displaced radars, prior information on target is available from existing high-definition maps and other sensor measurements. 
For instance, the position of target may follow a known normal distribution $\Phi\sim \mathcal{CN}(\Phi_0,\mathbf{R}_{\Phi})$. This prior knowledge is difficult to be incorporated into the general formulation of (\ref{eq:nonSPH}). Hence, instead, we consider the prior on $\mathbf{c}$ for which a common prior model is the Laplace distribution $p(\mathbf{c})\propto e^{-|\mathbf{c}|_1/\nu}$. Then, using the posterior distribution,
\begin{equation}
p(\mathbf{c}|\mathbf{r}_{nc},\mathbf{\alpha})=p(\mathbf{r}_{nc}|\mathbf{c},\mathbf{\alpha})p(\mathbf{c}),
\label{eq:postd}
\end{equation}
 and assuming the likelihood distribution to be Gaussian, the maximum \textit{a posteriori} probability (MAP) estimate leads to
\begin{equation}
J=|| \mathbf{r}_{nc}-\mathbf{H}_{\alpha}\mathbf{c}||^2+\mu ||\mathbf{c}||_1,
\label{eq:map}
\end{equation} 
where the regulation factor $\mu$ needs to be carefully selected to achieve good performance. It follows from (\ref{eq:map}) that the posterior distribution leads to a sparse imaging problem with unknown variables $\mathbf{\alpha}$. 
\subsubsection{Block-wise Orthogonal Matching Pursuit}
The conventional $\ell_1$-norm minimization does not exploit the block-sparse structure. We propose a block-wise Orthogonal Matching Pursuit (OMP) which is more efficient in implementation. 
We first ascertain the index of the potential target so that the correlation between $\mathbf{r}_k$ and the basis function $\mathbf{H}_s(\Phi_l)$ is maximized. This is equivalent to finding
\begin{equation}
\lambda_t=\text{argmin}_l ||r_k-\mathbf{H}_s(\Phi_l)\mathbf{\alpha}||_2.
\end{equation}
Once this block index is identified, find the coefficient so that the new residual is
\begin{equation}
\text{min}||r_k-\sum_{i\in \mathcal{S}_I}\mathbf{H}_s(\Phi_i)\bm{\alpha}_i||_2
\end{equation}
\subsubsection{Dimensionality issues}
One major concern in the above-mentioned imaging is processing the high-dimensional signal $\mathbf{z}_{nc}$. When the field of view is large, then the computation load becomes prohibitive. It is possible to mitigate this by employing CS-based sensing in the space and time domains \cite{mishra2019sub}. From (\ref{Eq:sigModCS}), assume the measurement matrix $\mathbf{G}$ is able to compress the measurements $\mathbf{r}_{nc}$ into a set of samples of smaller dimensions. 
The resulting signal is\par\noindent\small
\begin{equation}
\begin{split}
\mathbf{v}&=\mathbf{G}\mathbf{r}_{nc}=\mathbf{G}\mathbf{H}\mathbf{b}+\tilde{\mathbf{n}}_{nc} \triangleq \tilde{\mathbf{G}}\mathbf{b}+\tilde{\mathbf{n}}_{nc}\\
\end{split}
\end{equation}\normalsize
The block-sparse vector $\mathbf{b}$ may now be obtained with fewer samples and lower computational power. 

The non-coherent imaging provides better angular resolution with multiple radar sensors than using a single sensor. Apart from lack of synchronization, the non-coherency may arise from factors such as antenna array calibration error and target fluctuation in spatial domain. The improvement in the angular resolution is mainly because of the triangulation effect from the displaced sensor geometry. This radar imaging system is more flexible and robust to system errors. 

\subsection{Coherent imaging}
Coherent imaging exploits the fact that the time synchronization error is constant in each frame. Hence, if this error is estimated, then the limited antenna element separation on vehicles implies that different sensors have identical views of the target. Meanwhile, antenna phase impact is also constant and calibrated in advance. Thus, for each target, we have $\tilde{\alpha}_q=\tilde{\alpha}$. 
We now apply the non-coherent imaging procedure and estimate the coarse image. From the image, pick a single strong and isolated target (one could also use multiple targets in this step). Then, the received signal for the single target is\par\noindent\small
\begin{equation}
    \mathbf{y}_c=(\mathbf{\alpha}_c\otimes\mathbf{I}_{MNKN_s\times 1})\circ \mathbf{h}(\bm{\Phi}),
    \label{eq:sigmodcoh}
\end{equation}\normalsize
where $\mathbf{\alpha}_c=[\tilde{\alpha},\tilde{\alpha}c_2,\cdots,\tilde{\alpha}c_Q]$. Using this strong target, construct the matched filter according to $\mathbf{h}(\bm{\Phi})$ with identical target reflection coefficients. Then, assuming Gaussian noise, the maximum likelihood approach estimates the target coefficient from the measurement data. 

For each sensor, we get $\hat{\alpha}_q,q=1,\cdots,Q$ outputs. The relationship between $\hat{\alpha}_1, \hat{\alpha}_q$ is simply $\hat{\alpha}_q=\hat{\alpha}_1c_q$. Thus, from this relationship and the definition of $c_q$, we estimate the time synchronization error $\hat{\sigma}_q$. If we select more targets, the estimates are averaged for improved accuracy. Once the synchronization error estimated, the signal in (\ref{eq:sigmodcoh}) becomes\par\noindent\small
\begin{equation}
\begin{split}
    \mathbf{y}_c&=(\mathbf{\alpha}_c\otimes\mathbf{I}_{MNKN_s\times 1})\circ \mathbf{h}(\bm{\Phi})\\
    &=\tilde{\alpha}\mathbf{h}_c(\bm{\Phi}),
    \end{split}
    \label{eq:sigmodcohv1}
\end{equation}\normalsize
where\par\noindent\small
\begin{equation}
   \mathbf{h}_c(\bm{\Phi})=([1,c_1,\cdots,c_Q]^T\otimes\mathbf{I}_{MNKN_s\times 1})\circ \mathbf{h}(\bm{\Phi}) .
   \label{eq:hc}
\end{equation}\normalsize
From (\ref{eq:hc}), construct the PTRF of the coherent processing as\par\noindent\small
\begin{equation}
    \begin{split}
    F_c(\Phi)&=(\tilde{\alpha}\mathbf{h}_c(\Phi))^H(\tilde{\alpha}\mathbf{h}_c(\Phi))\\
    &=|\tilde{\alpha}|^2\mathbf{h}^H(\Phi)\mathbf{h}(\Phi).
    \end{split}
    \label{eq:CP-PTRF}
\end{equation}\normalsize
It follows from (\ref{eq:CP-PTRF}) that the integration gain is achieved over the column vector $\mathbf{h}(\Phi)$ and the resolution is limited now by the full aperture determined by all the individual radars. 

The coherent received signal with synchronization is
\begin{equation}
    \mathbf{r}_c=\mathbf{H}\mathbf{\alpha}+\mathbf{n}_c,
    \label{eq:SigModCohsynch}
\end{equation}
where $\mathbf{H}=[\mathbf{h}_c(\Phi_{1}),\cdots,\mathbf{h}_c(\Phi_{L})]$ is the basis matrix and $\mathbf{\alpha}=[\alpha_1;\cdots;\alpha_L]$ is a sparse vector with $\alpha_l=0$ if there is no target at $\Phi_l$ and $\alpha_l=\alpha_p$ only if the $p-$th targets is located at $\Phi_l$. Since the basis matrix is constructed with synchronization estimates $\mathbf{h}_c(\bm{\phi})$, this coherent imaging is simply a sparse recovery problem without block structure (different from non-coherent imaging) and is solved by conventional known methods \cite{foucart2013invitation}. With our proposed coherent imaging method, displaced radar sensors achieve the best angular resolution along with improved accuracy using non-coherent measurements from multiple radar sensors. 
\subsection{Bayesian CS Imaging}
The aforementioned non-coherent and coherent imaging methods are based on the deterministic CS, which does not exploit the prior information on the objects to be imaged. 
We incorporate Bayesian imaging by adopting relevance vector machine (RVM) which balances the efficiency in computations and the accuracy in the results. We briefly summarize the key steps of this algorithm; for more details, the reader may refer to \cite{tippingFastMLM2003,tippingSBLRVM2001,Ji2008BCS}. In RVM, a hierarchical prior is used. 
Assume the target distribution vector $\mathbf{\alpha}$ in (\ref{eq:SigModCohsynch}) is conditioned on the priors as
\begin{equation}
    p(\mathbf{\alpha}|\mathbf{\beta})=\prod_{i=1}^{L}\mathcal{CN}(\alpha_i|0,\beta_i^{-1}),
\end{equation}
where $\beta_i$ is the inverse-variance of a Gaussian density function. Then, a gamma prior is considered for $\mathbf{\beta}$, i.e.
\begin{equation}
    p(\mathbf{\beta}|a_1,b_1)=\prod_{i=1}^{L}\Gamma(\beta_i|a_1,b_1),
\end{equation}
as well as the inverse of the noise variance, i.e., $\beta_0=1/\sigma^2$, with
\begin{equation}
    p(\beta_0|a_2,b_2)=\Gamma(\beta_0|a_2,b_2),
\end{equation}
where 
\begin{equation}
    \Gamma(x|a_0,b_0)=\frac{(b_0)^{a_0}}{\Gamma(a_0)}x^{a_0-1}e^{-b_0x},
\end{equation}
with $\Gamma(x)$ being the gamma function.

The Bayesian recovery relies on the posterior probability distribution
$p(\mathbf{\alpha},\mathbf{\beta},\beta_0|\mathbf{r}_c)$,
which is decomposed as
\begin{equation}
    p(\mathbf{\alpha},\mathbf{\beta},\beta_0|\mathbf{r}_c)=p(\mathbf{\alpha}|\mathbf{r}_c,\mathbf{\beta},\beta_0)p(\mathbf{\beta},\beta_0|\mathbf{r}_c).
\end{equation}
The posterior distribution over the $\mathbf{\alpha}$ is 
\begin{equation}
    p(\mathbf{\alpha}|\mathbf{r}_c,\mathbf{\beta},\beta_0)=\mathcal{CN}(\mathbf{\alpha}|\mathbf{\mu},\Sigma),
\end{equation}
where the mean and covariance are, respectively,
\begin{align}
    \mathbf{\mu}&=\sigma^{-2}\Sigma\mathbf{H}^H\mathbf{r}_c, \\
    \mathbf{\Sigma}&=(\sigma^{-2}\mathbf{H}^H\mathbf{H}+\mathbf{B})^{-1},
\end{align}
with $\mathbf{B}=diag(\beta_1,\cdots,\beta_L)$. This means that in the presence of parameters $\mathbf{\beta},\beta_0$, the target scattering vector is inferred from the posterior distribution. Thus, the learning problem turns into the estimation of these parameters. In RVM, these are obtained from the data by performing the Type-II maximum likelihood procedure, i.e. to maximize the following over the parameters,
\begin{equation}
\begin{split}
    p(\mathbf{r}_c|\mathbf{\beta},\beta_0)&=\int p(\mathbf{r}_c|\mathbf{\alpha},\beta_0)p(\mathbf{\alpha}|\mathbf{\beta})d\mathbf{\alpha}\\
    &=\mathcal{Nc}(\mathbf{r}_c|0,C),
    \end{split}
\end{equation}
where $\mathbf{C}=\sigma^2\mathbf{I}+\mathbf{H}\mathbf{B}^{-1}\mathbf{H}^H$. The maximization of $p(\mathbf{\beta},\beta_0|\mathbf{r}_c)\propto p(\mathbf{r}_c|\mathbf{\beta},\sigma^2)p(\mathbf{\beta})p(\sigma^2)$ is equivalent to the maximization of $p(\mathbf{r}_c|\mathbf{\beta},\beta_0)$. 
The estimated $\mathbf{\beta}$ is
\begin{equation}
    \beta_i^{(\textrm{new})}=\frac{\gamma_i}{\mu_i^2},
\end{equation}
where $\mu_i=\mathbf{\mu}(i)$, $\gamma_i=1-\beta_i\Sigma(i,i)$ and noise variance estimate
\begin{equation}
    (\sigma^2)^{(\textrm{new})}=\frac{||\mathbf{r}_c-\mathbf{H}\mathbf{\mu}||^2}{L-\sum_i\gamma_i}.
\end{equation}

\section{Numerical Experiments}
\label{sec:numexp}
\begin{figure}[t]
    \centering
    \includegraphics[width=1.0\columnwidth]{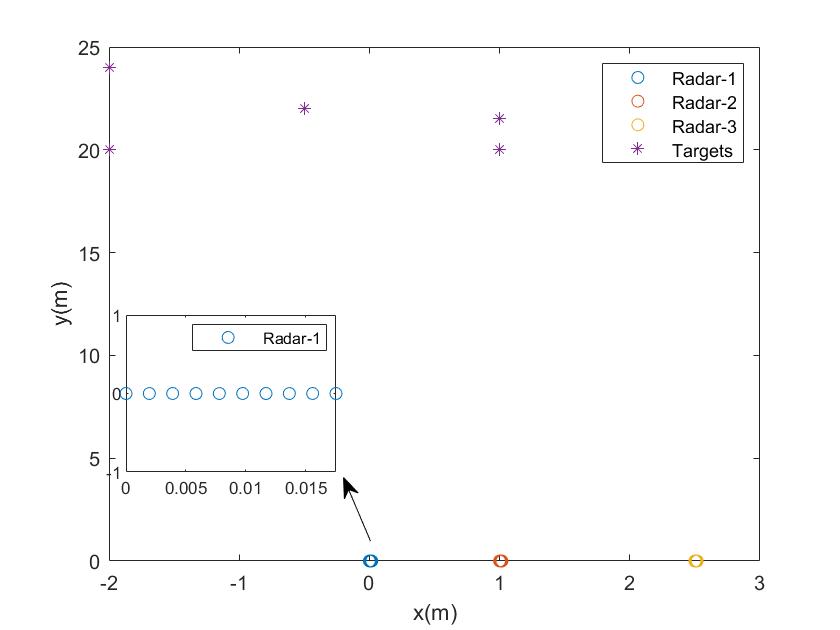}
    \caption{Illustration of displaced radars and target geometry.}
    \label{fig:RadarGeo}
\end{figure}

We validated our approach through extensive numerical experiments. In all simulations, we consider a front-looking radar configuration, which is more challenging in obtaining high resolution imaging. The geometry between radar and targets are illustrated in Fig.\ref{fig:RadarGeo}, where each of the radar has a virtual uniform linear antenna array. 
The general system setup and parameters are given in Table-\ref{tab:SimuParaSets}. In different simulation scenarios, there are some minor changes to the parameters which will be stated separately.
\subsection{Point target response function}
\label{subsec:ptrf}
\begin{figure}[t]
    \centering
    \includegraphics[width=1.0\columnwidth]{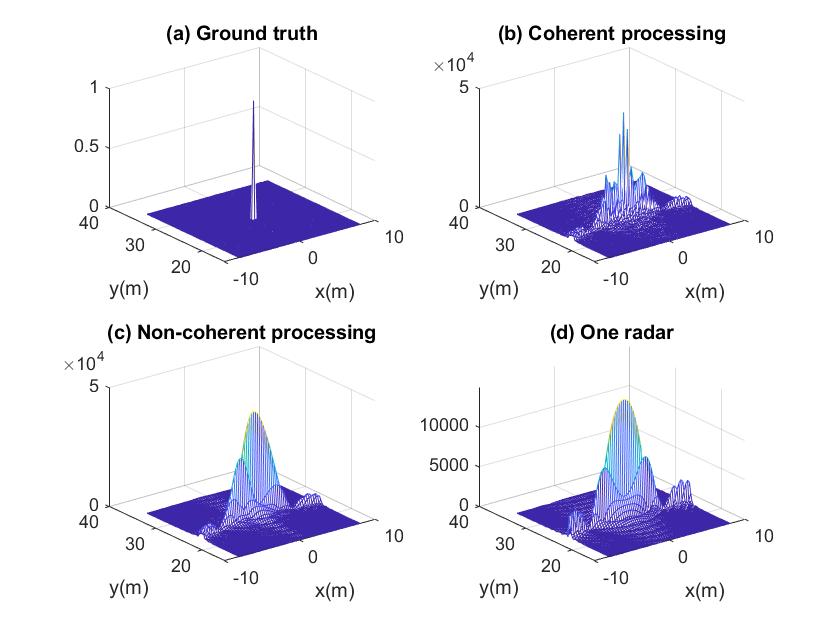}
    \caption{Comparison of PTRFs for various imaging methods.}
    \label{fig:PTRFcompM}
\end{figure}

\begin{figure}[t]
    \centering
    \includegraphics[width=1.0\columnwidth]{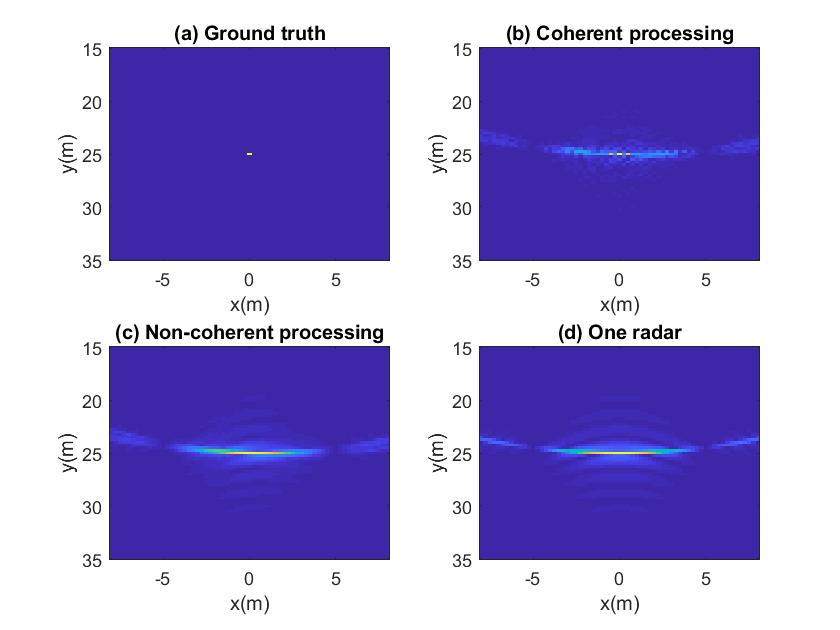}
    \caption{PTRF comparison with image plots.}
    \label{fig:PTRFcompI}
\end{figure}

\begin{figure}[t]
    \centering
    \includegraphics[width=1.0\columnwidth]{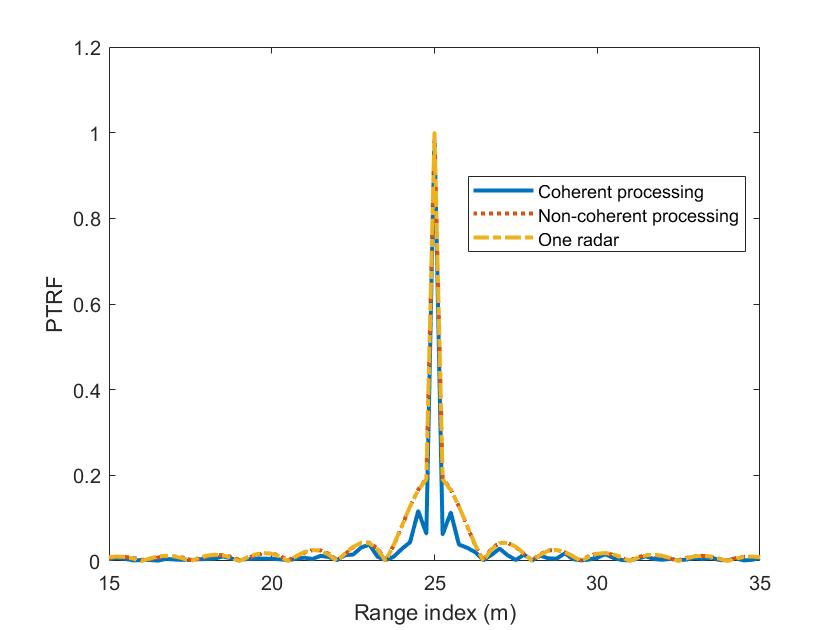}
    \caption{Range cuts of PTRFs.}
    \label{fig:PTRFcompRc}
\end{figure}

\begin{figure}[t]
    \centering
    \includegraphics[width=1.0\columnwidth]{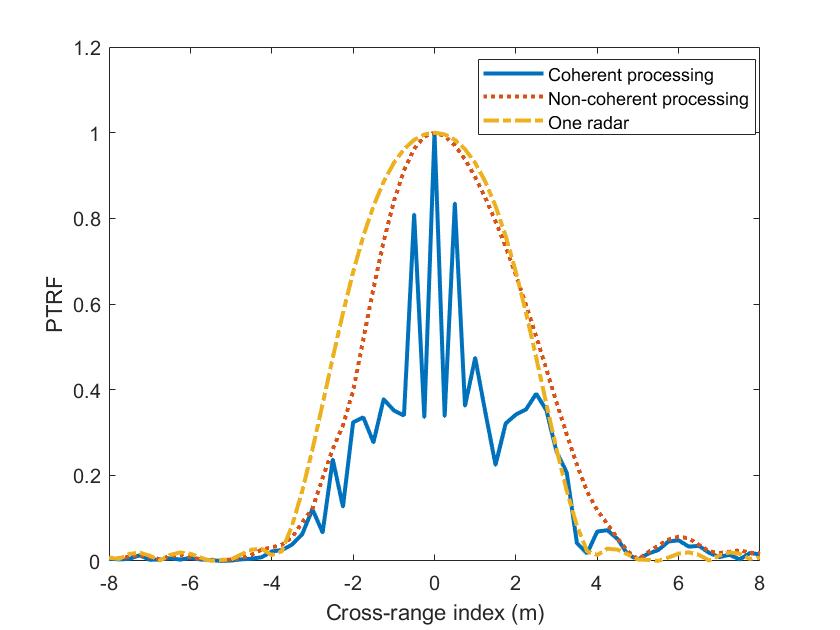}
    \caption{Cross-range cut of PTRFs.}
    \label{fig:PTRFcompCRc}
\end{figure}

In this section, we show the point target response function in a typical automotive radar scenario for single radar sensor, non-coherent processing, and processing in order to illustrate the properties, the potentials and challenges of various processing method. In the simulation, we take the following setup as listed in \ref{tab:SimuParaSets}.
\begin{table}[t]
\caption{Radar Simulation Parameters}
    \centering
\begin{tabular}{ |p{3cm}|p{3cm}|  }
 \hline
 \multicolumn{2}{|c|}{Radar parameters} \\
 \hline
 Radar sensor position &Radar velocity\\
 \hline
 Radar-1:(0m,0m) & (1,15)m/s\\
 Radar-2:(0m,1m) & \\
 Radar-3:(0m,2.5m) &  \\
 \hline
 \multicolumn{2}{|c|}{Target parameters} \\
 \hline
 Imaging area &Target position\\
 \hline
 X-area: [-8,8] & (0m,25m)\\
 Y-area: [15,35] & \\
 \hline
 \multicolumn{2}{|c|}{Radar waveform} \\
 \hline
 Bandwidth & 500MHz\\
 \hline
 $T_p$ & 5$us$\\
 \hline
 $f_s$ & 30MHz \\
 \hline
 $T_r$ & 30$us$ \\
 \hline
 $f_c$& 77:0.5:78 GHz\\
 \hline
 $K$& 10\\
 \hline
\end{tabular}
\label{tab:SimuParaSets}
\end{table}
We show the PTRF of single radar sensor, the non-coherent processing, and the coherent processing in Fig. \ref{fig:PTRFcompM} and Fig. \ref{fig:PTRFcompI}. From these two plots we can see that the single sensor enjoys low sidelobe level in the response function but with the worst resolution capability while the coherent processing has the highest sidelobe level and best resolution. The presence of the high sidelobe may disturb the imaging quality significantly in the coherent processing if not treated properly. The single radar processing has lower integration gain as compared to the non-coherent processing and coherent processing since it has only data from one radar sensor. Also looking into Fig. \ref{fig:PTRFcompRc} and Fig. \ref{fig:PTRFcompCRc}, the three method has similar resolution in the range dimension which is mainly determined by the bandwidth of radar signal. Meanwhile, the resolution of non-coherent processing in the cross-range dimension is slightly better than that of the single radar, thanks to the displacement between radars providing triangulation effect. In general, the resolution improvement is radar geometry dependent and better resolution can be achieved through optimizing the position of different radar sensors. Based on the point target response function, we can see that the major challenge in the coherent processing is the sidelobe suppression.

\subsection{Performance bounds}
\label{subsec:num_exp_perf}
\begin{figure}[t]
    \centering
    \includegraphics[width=1.0\columnwidth]{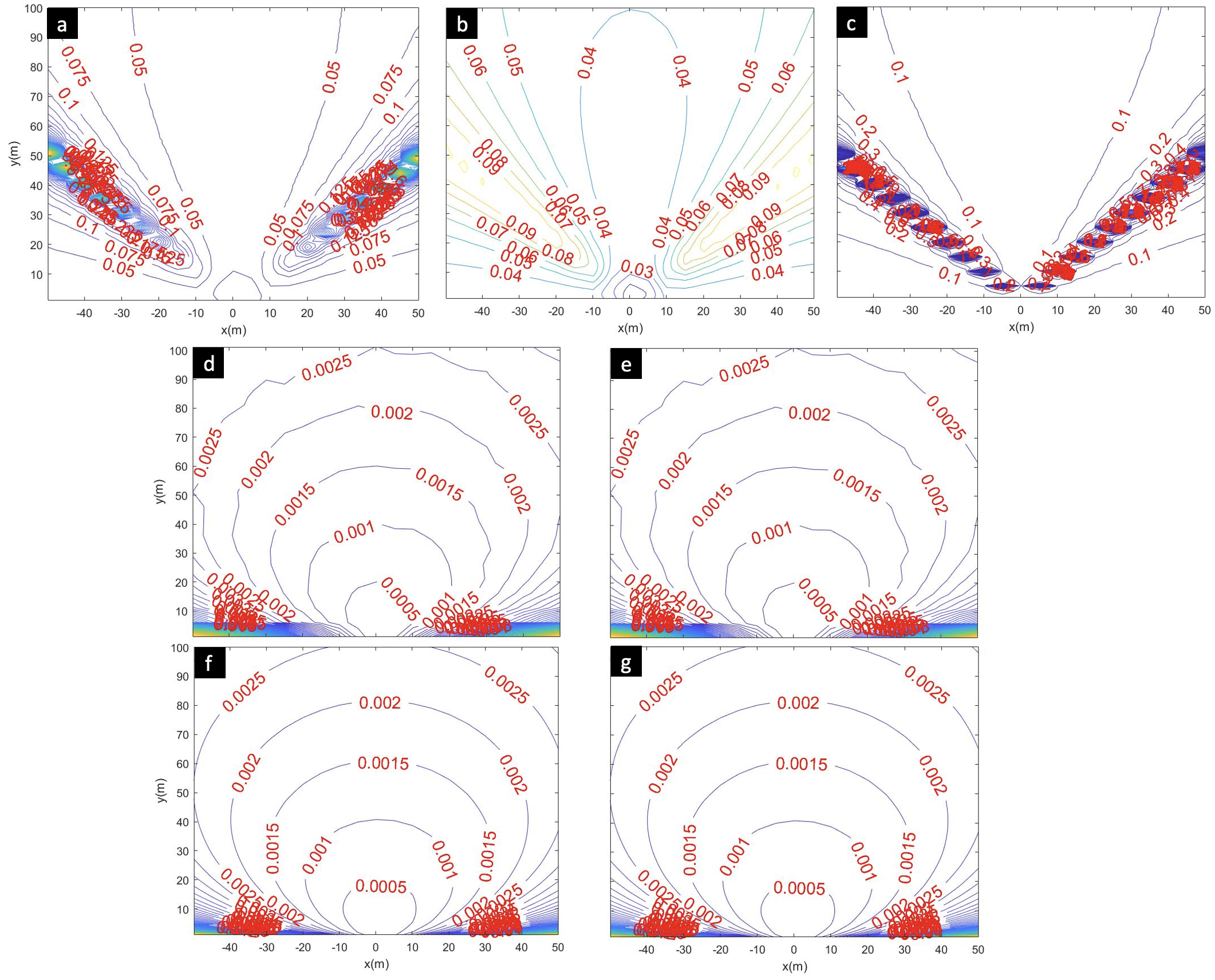}
    \caption{Average bound on the position, i.e., $\bar{\mathbf{F}}_{\bm{\Phi}}$. (a) Point-cloud fusion CRLB with $Q=3$ radars, (b) Point-cloud fusion BCRLB, $Q=3$, (c) Point-cloud fusion CRLB, $Q=1$, (d) Non-coherent processing CRLB, $Q=3$, (e) Non-coherent processing BCRLB, $Q=3$, (f) Coherent processing CRLB, $Q=3$, and (g) Coherent processing BCRLB, $Q=3$. The radars are located at $[0,0]$ m, $[1,0]$ m, and $[2,0]$ m. The $Q=1$ case corresponds to the first radar.}
    \label{fig:BCRB}
\end{figure}
We evaluated different performance bounds on the accuracy of position through numerical experiments. In particular, we compute the average of the bounds on positions, i.e.\par\noindent\small
\begin{align}
 \bar{\mathbf{F}}_{\bm{\Phi}}=\frac{\mathbf{F}_{\bm{\Phi}}^{-1}(x,x) +\mathbf{F}_{\bm{\Phi}}^{-1}(y,y)}{2}.   
\end{align}\normalsize
In all experiments, each MIMO radar has two transmitters and four receivers. The receive and transmit antennas are spaced at half-wavelength and two wavelengths, respectively. The center frequency is $77$ GHz with signal bandwidth $150$ MHz, chirp duration $5$ $\mu$s and sampling frequency $10$ MHz. We consider $Q=3$ radar located at coordinates $[0,0]$ m, $[1,0]$ m, and $[2,0]$ m within the \textit{ego-car} coordinate system. 
The input SNR is -30 dB. Based on the CRLBs of frequency and DoA, this SNR leads to range estimation accuracy of $0.06$ m, and DoA estimation accuracy of $0.02^{\circ}$ (at $45^{\circ}$ DoA). 

Using only one radar sensor, the point-cloud fusion provides a location accuracy of about $0.1$ m. The prior information on the target follows a Gaussian distribution centered at the true position and $0.1$ m standard deviation. The target area of interest ranges $[-50,50]$ m and $[0,100]$ m in x- and y-dimensions. We numerically computed BCRLB conditions on the prior over 20 Monte-Carlo trials. Figure~\ref{fig:BCRB}a and \ref{fig:BCRB}b show contours of the CRLB and BCRLB for point-cloud fusion using three radars while Fig.~\ref{fig:BCRB}c plots CRLB of one radar with respect to the distance from radar sensors. We note that, when conventional processing is used without considering prior information, using three sensors significantly improves the performance of positioning accuracy over a single radar. Exploiting prior information provides additional improvement. Figure~\ref{fig:BCRB}e illustrates BCRLB for, \textit{ceteris paribus}, non-coherent processing with $Q=3$ sensors. It is clear that non-coherent mode significantly improves the accuracy of multiple displaced sensors. We further observed that the coherent processing outperforms the non-coherent mode by a very small margin (Figs.~\ref{fig:BCRB}d-g).
\subsection{Non-coherent imaging}
We now illustrate the imaging performance of the single radar sensor, and non-coherent imaging algorithms with and without exploiting the block sparsity structure using various sparsity recovery algorithms. We consider a medium range imaging scenario with 5 targets located at (-2, 20)m, (-2, 24)m, (-0.5, 22)m, (1,20)m, and (1, 21.5)m. As for the recovery algorithm, we consider $\ell_1$-norm minimization, OMP and block OMP algorithm. The non-coherent imaging performance is illustrated in the Figs. \ref{fig:BOMPcomp}, \ref{fig:BOMPcomp2}, \ref{fig:BOMPcomp3} below. From these figure we can see that Using only one sensor in general leads to worse performance than the non-coherent processing with 3 radar sensor. Meanwhile, L1-norm can outperform the OMP if the noise level is properly tuned. However, its computation is much slower than the OMP approach. We thus choose not to evaluate the non-coherent processing using L1-Norm since the computation time is even longer. Regarding the non-coherent processing, with block sparsity constraint can lead to improved imaging quality as compared with exploiting sparsity directly. In particular, from Fig. \ref{fig:BOMPcomp2}, we see that with sparsity constraint, the recovered targets have the same structure over the 3 radar sensors. On the contrary, directly exploiting sparsity may lead to different image structure over the 3 radars. In addition, among all the targets, there are two with the same y-axis and different x-axis. Those two targets roughly located in the same range cell of the first radar sensor and the separation is within the angular resolution. Thus, it will be more difficult for the first sensor to reliably recover both with high quality. The block sparsity recovery based non-coherent processing can exploit the geometry advantage to have better resolution and imaging instead.
\begin{figure}[t]
    \centering
    \includegraphics[width=1.0\columnwidth]{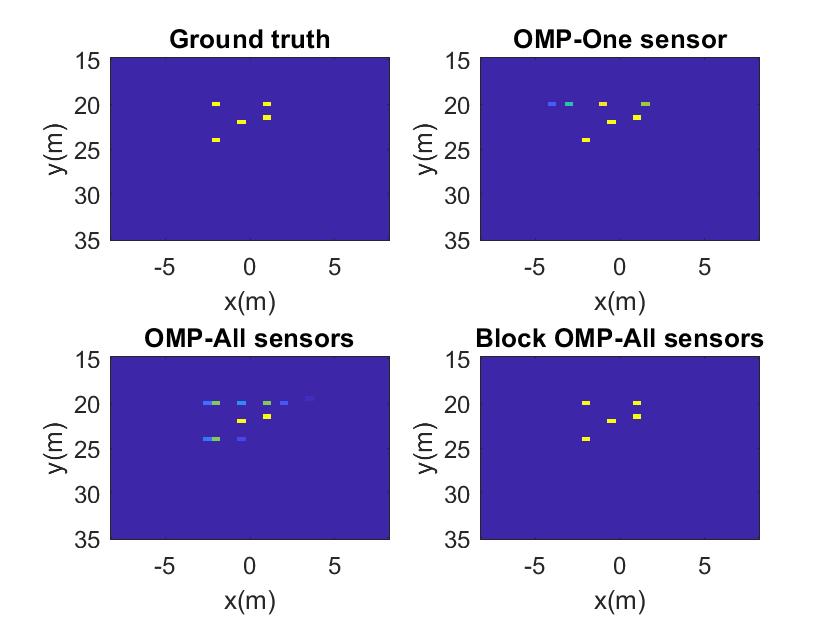}
    \caption{Performance comparison of BOMP and OMP for non-coherent imaging}
    \label{fig:BOMPcomp}
\end{figure}
\begin{figure}[t]
    \centering
    \includegraphics[width=1.0\columnwidth]{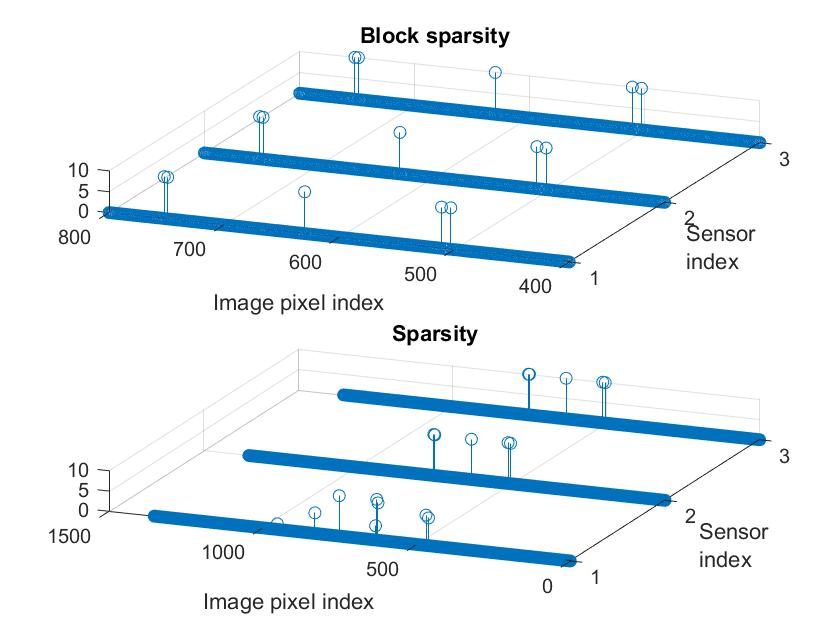}
    \caption{Performance comparison of BOMP and OMP for non-coherent imaging.}
    \label{fig:BOMPcomp2}
\end{figure}
\begin{figure}[t]
    \centering
    \includegraphics[width=1.0\columnwidth]{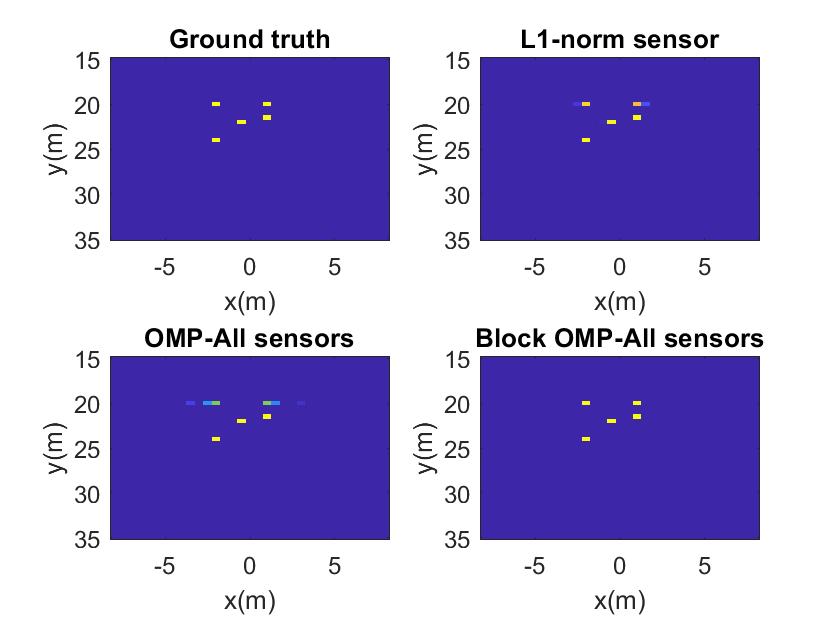}
    \caption{Performance comparison of BOMP, OMP, and $\ell_1$-N=norm for non-coherent imaging.}
    \label{fig:BOMPcomp3}
\end{figure}
\subsection{Coherent imaging}
\begin{figure}[t]
    \centering
    \includegraphics[width=1.0\columnwidth]{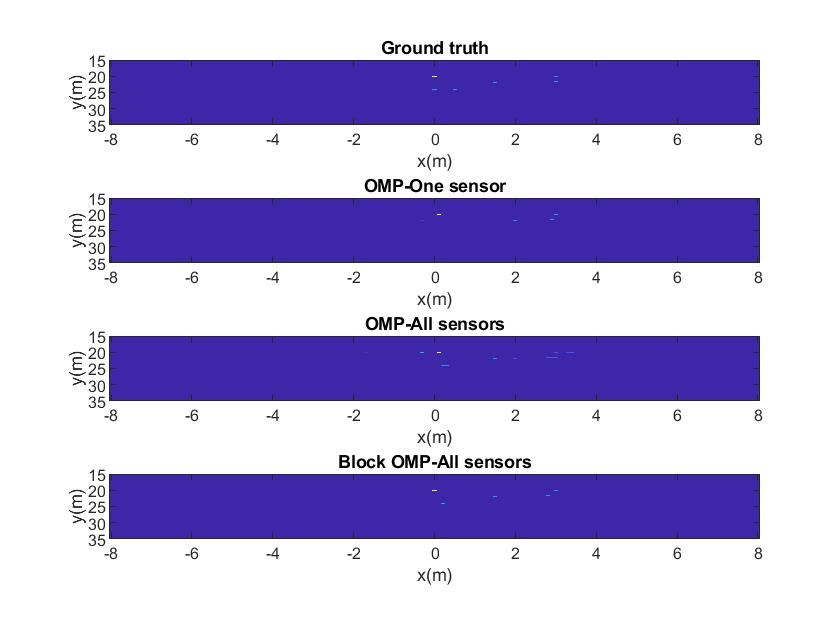}
    \caption{Non-coherent imaging with closely spaced targets.}
    \label{fig:sim4NCP}
\end{figure}

\begin{figure}[t]
    \centering
    \includegraphics[width=1.0\columnwidth]{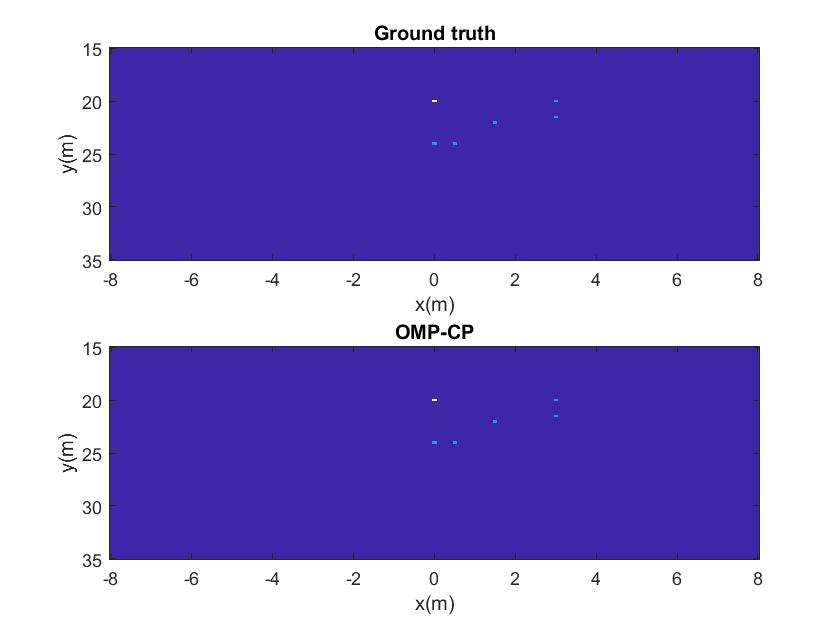}
    \caption{Coherent imaging with closely spaced targets.}
    \label{fig:sim4CP}
\end{figure}

\begin{figure}[t]
    \centering
    \includegraphics[width=1.0\columnwidth]{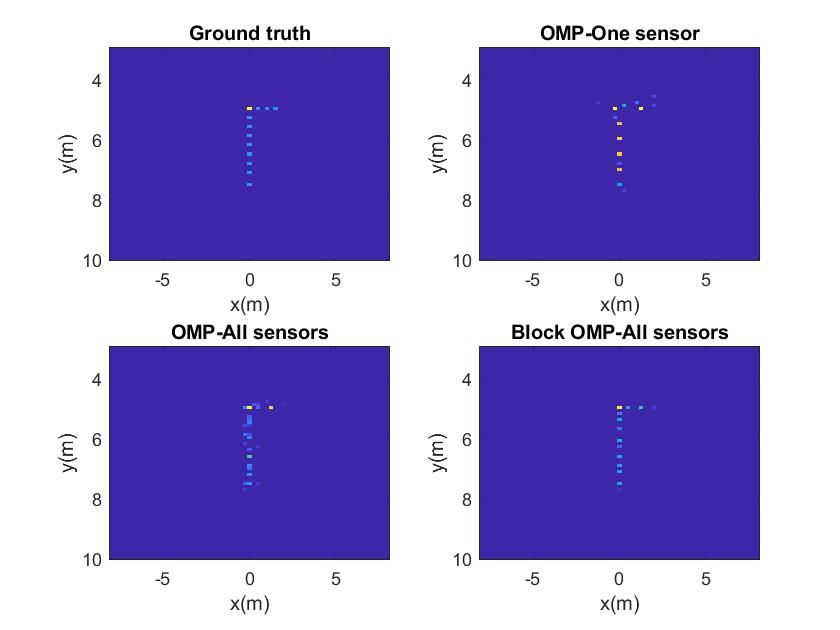}
    \caption{Non-coherent imaging of near range targets.}
    \label{fig:NearRangeNCP}
\end{figure}

\begin{figure}[t]
    \centering
    \includegraphics[width=1.0\columnwidth]{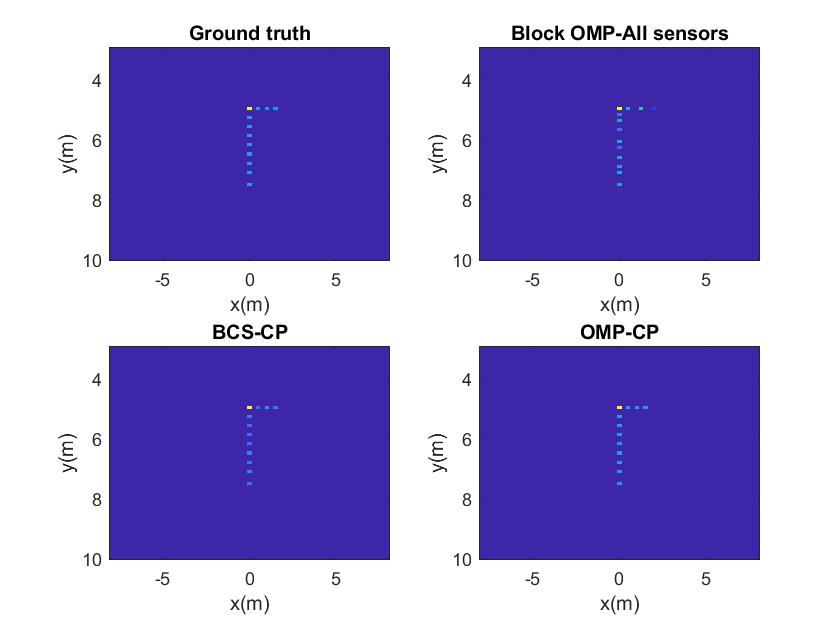}
    \caption{Coherent imaging of near range targets.}
    \label{fig:NearRangeCP}
\end{figure}

\begin{figure}[t]
    \centering
    \includegraphics[width=1.0\columnwidth]{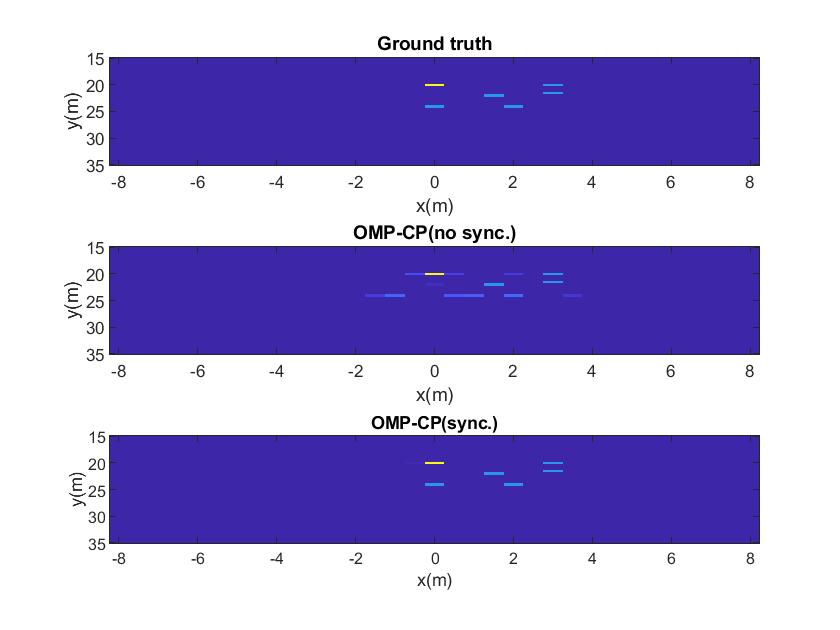}
    \caption{Time offset effect on coherent imaging.}
    \label{fig:sim4CPTOeffect}
\end{figure}

Simulation results is given here to illustrate the improved performance of the proposed coherent imaging. In this simulation, we consider the case there are two closely spaced targets located at (0m, 24m) and (0.5m, 24m) among a group of 5 targets. They cannot be separated by the non-coherent processing, and the typical recovery results can be seen from Fig. \ref{fig:sim4NCP}. However, after processing based on the proposed coherent imaging, the two targets can be clearly seen in the image, as can be seen from typical recovery results like that in Fig. \ref{fig:sim4CP}. The target at the most top-left corner is selected for synchronization estimation. The synchronization error in this simulation is set as 10us and 5us for sensor-2 and sensor-3 with respect to sensor-1, respectively. From these results, we can see that, with automatic synchronization the coherent imaging can achieve better resolution than the non-coherent imaging. The impact of time offset on the imaging is shown in Fig. \ref{fig:sim4CPTOeffect}. We can see that the presence of time offset if not properly handled, would blur the image of coherent imaging. We further consider some near range imaging scenario. In this scenario, there are 4 reflection points closely spaced with each other within the angular resolution with the same range and there are 9 points closely space with each other along the range direction with distance of one range resolution determined by the bandwidth of individual radar. In the target geometry configuration, there are four points with the same coordinate in y-axis of 5 meter and spaced with equal distance of 0.5 meter in x-axis. Meanwhile, there are 8 points with same x-axis coordinate of 0 meter and equally spaced in y-axis with distance of 0.3 meter. Thus, unlike the coherent imaging method, the non-coherent imaging could not recover all the points with their actual positions. The coherent processing can achieve good recovery for both target position and coefficient. The results are shown in Fig. \ref{fig:NearRangeNCP} and Fig. \ref{fig:NearRangeCP}. 

\subsection{Bayesian imaging}
\begin{figure}[t]
    \centering
    \includegraphics[width=1.0\columnwidth]{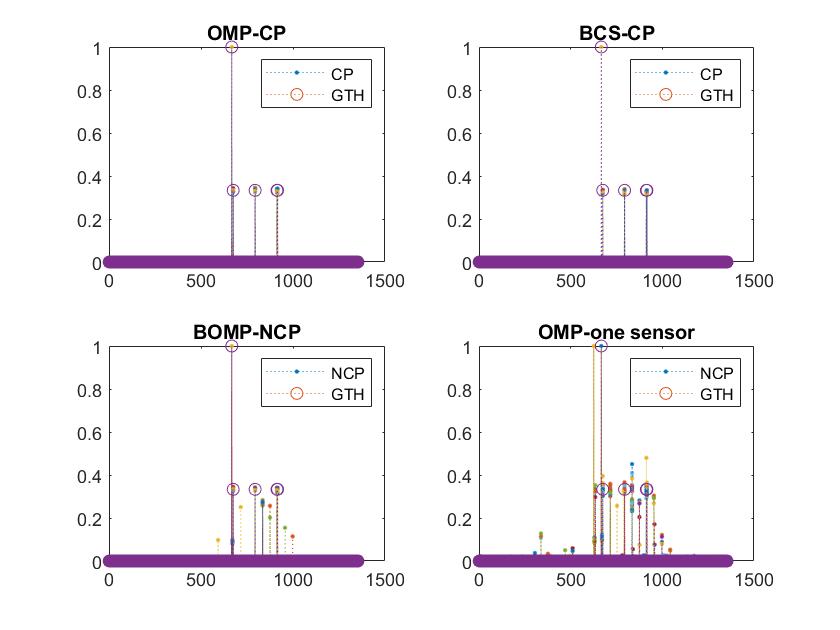}
    \caption{Recovery of different approaches over 100 trials.}
    \label{fig:mmseOvlp}
\end{figure}

We use the Bayesian CS approach together with the automatic synchronization. It is implemented based on the approach in \cite{Ji2008BCS} by modifying the implementation for real valued data with complex valued data. Based on the experimental evaluation, as compared with coherent imaging, Bayesian approach can provide very similar yet more reliable image recovery in higher SNR than the OMP based coherent imaging. However, in low SNR, Bayesian approach tends to give more false alarms although the target recovery remains good. Meanwhile, the time offset has similar impact on the Bayesian imaging as on coherent processing using OMP recovery, which showing the blurring of the image due to the synchronization error. 
In order to compare the different imaging schemes more precisely, we provide some quantitative measuring of the imaging quality of each of the proposed imaging algorithms. In this case, we measure the normalized mean square error(NMSE) of the target coefficient estimation for comparison, which is defined as,
\begin{equation}
    NMSE_{All}=||\mathbf{\alpha}/|\mathbf{\alpha}|_{max}-\hat{\mathbf{\alpha}}/|\hat{\mathbf{\alpha}}|_{max}||_2
    \label{eq:NMSE}
\end{equation}
and the NMSE for targets only:
\begin{equation}
    NMSE_{target}=||\mathbf{\alpha}_t/|\mathbf{\alpha}|_{max}-\hat{\mathbf{\alpha}}_t/|\hat{\mathbf{\alpha}}|_{max}||_2
    \label{eq:NMSEtgt}
\end{equation}
The reason for evaluating only the target NMSE is that the recovered imaged sometime has false alarms scattered around with much smaller magnitude which can be eliminated in the post-processing stage using detection techniques. Thus, this metric may better reflect the actual accuracy of target imaging. In this study, we evaluate the NMSE versus the signal-to-noise ratio for different imaging schemes. 100 Mont Carlo trials are run to generate the results. The target scenario is similar with the one used in non-coherent imaging but that the first target is 7dB higher than the rest of the target. The results are shown in Table-\ref{tab:NMSE} below.
\begin{table}[t]
    \caption{NMSE comparison of different imaging schemes}
    \centering
    \begin{tabular}{c|c|c|c|c|c}
        \hline
         SNR & NMSE & One sensor & BOMP-NCP & OMP-CP & BCS-CP \\
         \hline
         -10dB & All &0.82 &0.67 &0.67 & 83 \\
               & Target &0.74 &0.67 &0.67 & 0.17 \\
         \hline
         -5dB & All &0.87 &0.62 &0.58 & 0.83 \\
              & Target &0.77 &0.60 &0.58 & 0.09\\
         \hline
         -0dB & All &1.01 &0.63 &0.58 & 0.35 \\
              & Target &0.81 &0.60 &0.58 & 0.04\\
         \hline
         5dB & All &0.88 &0.36 &0.36 & 0.16 \\
              & Target &0.74 &0.35 &0.36 & 0.03\\
         \hline
         10dB & All &0.63 &0.19 &0.02 & 0.02 \\
              & Target &0.49 &0.15 &0.01 & 0.01\\
         \hline
    \end{tabular}
    \label{tab:NMSE}
\end{table}
We also provide the vectorized target scattering coefficient recovered by the algorithms over the 100 Mont Carlo trials to illustrate the performance. It is shownn in Figure.\ref{fig:mmseOvlp}. 
Meanwhile, computation time wise, the OMP based implementation is far more efficient than the BCS, not mentioning about the L1-Norm methods. So OMP based methods based on this study is a better choice considering both performance and computation time.
\section{Summary}
\label{sec:summ}
We studied the automotive radar imaging with displaced sensors that does not require accurate synchronization. We first derived the signal model of the asynchronous displace radar sensor system, based on which three imaging schemes were then proposed. The imaging performance of these three imaging schemes were investigated according to the BCRLB. The performance bounds suggested that the non-coherent imaging and coherent imaging can improve the accuracy of the displaced radar imaging significantly as compared with single radar sensor and point-cloud fusion using multiple radars. These results motivated the development of non-coherent imaging and coherent imaging. In particular, the non-coherent imaging has been formulated as a block-sparsity recovery problem while the coherent imaging is developed on the top of the automatic synchronization scheme. Bayesian approach to the coherent imaging was also studied that is able to exploit the prior information and shows improved performance in certain scenarios. As compared to non-coherent imaging, coherent imaging can achieve both improved accuracy and improved resolution. Simulation studies were conducted and the results show that the displaced sensor imaging can be one promising system to improve both the resolution and accuracy of automotive radars. 

\bibliographystyle{IEEEtran}
\bibliography{main}
\end{document}